\documentclass[11pt,a4paper]{article}

\usepackage[dvipsnames]{xcolor}
\pagecolor{white}

\usepackage{geometry}
\usepackage[utf8]{inputenc}
\usepackage{comment}
\usepackage{amsmath}
\usepackage{amsfonts}
\usepackage{amssymb}
\usepackage[final]{graphicx}
\usepackage{cite}
\usepackage{hyperref} 
\usepackage{cleveref}

\hypersetup{colorlinks,bookmarksopen,bookmarksnumbered,
citecolor=red,
linkcolor=red,
pdfstartview=false,
urlcolor=blue}

\newcommand{\ket}[1]{|#1\rangle}

\newcommand{\erf}[1]{\eqref{#1}}
\renewcommand{\th}{\vartheta}
\newcommand{\Th}{\Theta}
\newcommand{\eps}{\varepsilon}

\begin{document}

\title{Confinement in the spectrum of a Heisenberg-Ising spin ladder}
\author{Gianluca Lagnese$^{1,2}$, Federica Maria Surace$^{1,3}$, M\'arton Kormos$^4$, Pasquale Calabrese$^{1,2,3}$\\
$^1${\small SISSA, via Bonomea 265, 34136 Trieste, Italy}\\
$^2${\small INFN, Sezione di Trieste, 34136 Trieste, Italy}\\
$^3${\small International Centre for Theoretical Physics (ICTP), Strada Costiera 11, 34151 Trieste, Italy}\\
$^4${\small MTA-BME Quantum Dynamics and Correlations Research Group,}\\
{\small Budapest University of Technology and Economics, Budafoki \'ut 8., 1111 Budapest, Hungary}
}

\date{}

		\maketitle

\begin{abstract}
The Heisenberg-Ising spin ladder is one of the few short-range models showing confinement of elementary excitations without the need of an external field, 
neither transverse nor longitudinal. This feature makes the model suitable for an experimental realization with ultracold atoms. 
In this paper, we combine analytic and numerical techniques to precisely characterize its spectrum in the regime of Hamiltonian parameters showing confinement. 
We find two kinds of particles, which we dub intrachain and interchain mesons, that correspond to bound states of kinks within the same chain or between different 
ones, respectively. 
The ultimate physical reasons leading to the existence of two families of mesons is a residual double degeneracy of the ground state: the two types
of mesons interpolate either between the same vacuum (intrachain) or between the two different ones (interchain). 
While the intrachain mesons can also be qualitatively assessed through an effective mean field description and were previously known, the interchain ones 
are new and they represent general features of spin ladders with confinement. 
 
\end{abstract}

\section{Introduction}
Confinement of elementary particles is renowned as a fundamental mechanism for our understanding of fundamental interactions of nature. 
The prototypical example of this phenomenon is quark confinement in quantum chromodynamics \cite{Wilson1974} which is a ruling principle of strong interactions: 
quarks cannot be isolated at low-energy and they are only found in composite particles called hadrons such as baryons and mesons.
Indeed, the mass of ordinary matter is in an overwhelming majority in the binding energy of protons and neutrons rather than in the masses of 
truly elementary particles such as quarks and electrons. 

Remarkably, confinement of excitations is a relevant phenomenon also in condensed matter physics, as theoretically proposed in the late 
seventies \cite{McCoy1978,Ishimura1980} and directly verified in the last decade in a number of experiments with inelastic neutron scattering or other spectroscopic probes 
\cite{Lake2009,Morris2014,Grenier2015,Wang2015b,Wang2016,Wang2018,Wang2019}. 
To date, confinement has been found and studied in great detail in many one-dimensional and quasi-one-dimensional magnetic insulators, with Ising-like \cite{McCoy1978,Ishimura1980,Rutkevich1999,Bhaseen2004,Rutkevich2005} or Heisenberg-like \cite{Hida1991,Shelton1996,Schulz1996,Essler1997,Sandvik, Nersesyan1997,affleck1998soliton,Augier1999,Greiter2002,Greiter2002a} interactions. 
In all these cases, the spin-1/2 excitations (kinks or spinons) not only form bound states with integer spin, as a consequence of an (even weak) attractive interaction, 
but they cannot be observed as free particles at low energy, exactly like quarks in high energy physics. 
Following the pioneering work by McCoy and Wu \cite{McCoy1978}, confinement has been studied with various analytical and numerical methods 
in thermal equilibrium  \cite{Delfino1996,DELFINO1998675,Fonseca,Jung2006, Rutkevich2008,Rutkevich2010,Steinigeweg2015,Bera2017,Tonegawa2018,Rutkevich2018,CiracGaussian,Suzuki2018,Brenes2018,Fan2020,PaiPretkoFractonsLGT}.
In very recent times, it has been proposed that many quantitative aspects of confinement (such as the masses of the bound states) 
can be accessed very effectively  following the non-equilibrium real time dynamics \cite{marton2017}, a protocol which is routinely exploited in
ultracold atoms and trapped ions experiments.
This observation started an intensive theoretical activity on the 
subject \cite{jkr-19,rjk-19,Liu2018,Verdel19_ResonantSB,ch2019confinement,mazza2019,lerose2019quasilocalized,Castro-Alvaredo2020,vk-20,mrw-17,cr-19,vcc-20}
that lead to direct experimental implementation of a quantum simulator with trapped ions \cite{tan2019observation},
as well as several new theoretical and experimental ideas to understand lattice gauge theories in real time \cite{Cai2012,mdfppe,Hebenstreit2013,Pichler:2016it,Martinez2016,Surace}.

In this work, we wish to study and characterize accurately the spectrum of yet another model displaying confinement of elementary excitations. 
This system consists of two XXZ spin-1/2 chains coupled in an anisotropic manner, along the longitudinal (easy axis) direction via an Ising-like coupling. 
Explicitly the Hamiltonian is given by 
\begin{equation}
\begin{split}
\label{eq:ladder}
	    H(\Delta_{||},\Delta_{\perp}) = &\frac{J}{2} \sum_{j=1}^{L} \left[ \sigma^x_{j,1} \sigma^{x}_{j+1,1} +
	    \sigma^{y}_{j,1} \sigma^{y}_{j+1,1} + \Delta_{||}(\sigma_{j,1}^{z} \sigma_{j+1,1}^{z} + 1)\right]   \\
	    +& \frac{J}{2} \sum_{j=1}^{L} \left[ \sigma^{x}_{j,2} \sigma^{x}_{j+1,2} +
	    \sigma^{y}_{j,2} \sigma^{y}_{j+1,2} + \Delta_{||}(\sigma_{j,2}^{z} \sigma_{j+1,2}^{z} + 1)\right]   \\
	    +& J\Delta_{\perp}\sum_{j=1}^{L} \sigma^{z}_{j,2} \sigma^{z}_{j,1}\,.
\end{split}
\end{equation}
Here $\sigma^\alpha_{j,k}$ denotes the Pauli spin operators at the $j^\text{th}$ site of chain $k\in\{1,2\}$ and we impose periodic boundary conditions, 
$\sigma^\alpha_{L+1,k}=\sigma^\alpha_{1,k}$. The parameter $J$ sets the energy scale, throughout the paper we set it to $J=1$.
We focus on the regime $\Delta_{||}  \in (1,+\infty)$
for the anisotropy parameter of the chains which corresponds to their gapped antiferromagnetic phases. 
The last term couples the two spin chains with an Ising-like interchain interaction. 
Without losing generality, we set $\Delta_\perp>0$, i.e. an antiferromagnetic coupling of the chains.  
The sign of $\Delta_\perp$ can be reversed with a spin flip applied to one of the two chains, without altering the spectrum.

The model is highly symmetric. 
The $z$ component of the total magnetization on each chain, $M_k=\sum_j\sigma^z_{j,k},$ is conserved, corresponding to two $U(1)$ symmetries. For $\Delta_{\perp} =0$ there is a $\mathbb{Z}_2\times \mathbb{Z}_2$ symmetry associated with the spin flip $\sigma_{j,k}^{y/z}\to-\sigma_{j,k}^{y/z}$ of each chain. The coupling $\Delta_\perp \neq 0$ explicitly breaks this symmetry: the residual one is the global spin flip of both chains (a single $\mathbb{Z}_2$). 
Moreover, there is an additional symmetry related to the swapping of the chains $\sigma^\alpha_{j,1}\leftrightarrow\sigma_{j,2}^\alpha$. Finally,  
due to translational invariance, the energy levels are organized as eigenstates of the total momentum $P$.

One of the motivations to study the effects of confinement in  antiferromagnetic ladders lies in the fact that,
in contrast to spin chains where confinement is triggered by 
a symmetry-breaking field or long-range interactions, 
in a ladder geometry the confining potential naturally emerges as the effect of the (even small) local interaction between the chains, 
as can be easily seen in a mean field treatment \cite{Bhaseen2004,Schulz1996,Essler1997,Sandvik}.
This mechanism is reminiscent of quark confinement in chromodynamics, where the role of the attractive potential is played by the gauge field, i.e. the mediator of strong interactions. 
Similarly, in a ladder, the spins of one chain are additional degrees of freedom which effectively mediate the interactions between the particles of the other chain.
Consequently, the external field is not required because the staggered magnetization of one chain provides an effective staggered field for the other.
There are various possible ladders featuring confinement (e.g., those composed of Ising-like chains), but many of these require an external magnetic field which imposes difficulties in prospective cold atomic realizations (see however \cite{svmt-11}).
 Here we focus instead on coupled Heisenberg-type spin chains described by the Hamiltonian 
\eqref{eq:ladder} 
in which no external field, either longitudinal or transverse, makes an appearance, making them suitable for cold atom experiments.  
For example, the two chains in Eq.~\eqref{eq:ladder} can be mapped by a Jordan--Wigner transformation to spinless fermions coupled by a density-density interaction. This model can be easily realized by freezing the spin degrees of freedom in real fermion gases, e.g. utilizing the techniques of Ref. \cite{lens} for Ytterbium atoms.
Alternatively, one can use the true spin degrees of freedom and freeze the charge degree of freedom in spin $1/2$ fermionic condensates.

There is however a fundamental difference between confinement in ladders and chains with an external field.
Indeed, since for the ladders one symmetry is spontaneously broken, there are (at least) two true ground states (vacua) and the neutral mesons with respect to the confined charge 
can interpolate between the same vacuum or between different ones. 
This feature leads to the existence of two types of neutral bound states which we dub intrachain (`Type 1') and interchain (`Type 2') mesons interpolating, respectively, 
between the same or different vacua.  
We stress that `Type 2' interchain mesons are not charged bound states that exist in some other theories and interpolate between true and false vacua. 

A recent work \cite{Rutkevich2018} studied the properties of bound states on an XXZ chain in the anti-ferromagnetic region with the confining potential provided 
by an external staggered field.
This model can be interpreted as arising from a mean field treatment of the XXZ ladder in \eqref{eq:ladder}, with the staggered field encoding the mean field effect of one chain on the other  \cite{Bhaseen2004,Schulz1996,Essler1997,Sandvik}. Although such a mean-field treatment is rather accurate to capture an entire family of bound states, it completely misses another one. 
We go beyond this approximation, and study the full system in the strong anisotropy regime. 
As generally anticipated above, the main new theoretical insight is that beyond the already known (intrachain) mesons that also appear in the mean field approach, 
we identify another class of bound states that we dub interchain mesons. 

The paper is structured in the following way. In Section \ref{sec:exc} we describe the elementary excitations (intrachain and interchain mesons) and study their spectrum in the strong anisotropy regime. In Section \ref{sec:BS} we use a semiclassical approach to find a more accurate estimate of the spectrum in the regime of moderate anisotropy. In Section \ref{sec:compexc} we introduce an approximation to capture the spectrum of two-meson states. In Section \ref{sec:nat} we discuss the nature of the first excited states and find a transition as a function of the anisotropy parameters. 
In Section \ref{concl} we draw our conclusions. 
In the Appendix we report some details about semiclassical quantization.

\begin{figure}[!t]
\centering
\includegraphics[width=0.82\textwidth]{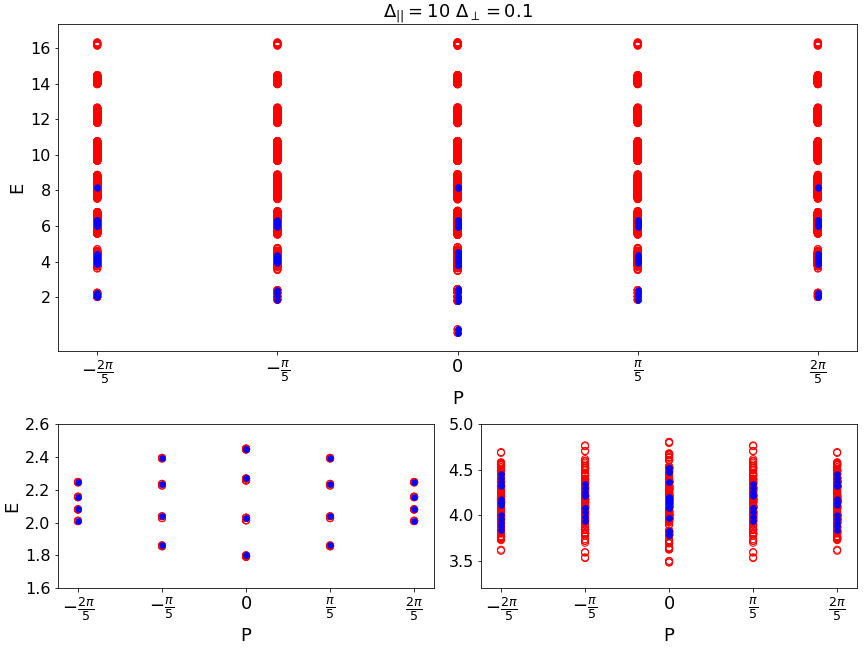}
\caption{Spectrum of the Heisenberg-Ising ladder Hamiltonian \eqref{eq:HI} for $\Delta_{||}=10$ and $\Delta_\perp=0.1$ (red circles)
compared with the XXZ-chain in a staggered field (blue symbols), see Eq.~\eqref{eq:HS}. Both spectra are in the sector of zero magnetization.
{\em Top:} Entire spectrum of both Hamiltonians for $L=10$. For large enough $\Delta_{||}$, the spectrum is organized in bands with fixed 
number of particles (kinks). The ladder has many more states than the corresponding chain.
{\em Bottom:} Zooms of the two-particle and four-particle sectors close to energy $E\sim 2$ and $E\sim4$ respectively. 
For the two-particle sector and even $L$, the spectra of the two models are in one-to-one correspondence (modulo a four-fold degeneracy which is resolved at higher order in $\epsilon$). 
This is no longer the case for the four-particle sector (and for odd $L$ in the two-particle sector). 
}
\label{fig:tot}
\end{figure}

\section{Elementary excitations in the strong anisotropy regime}
\label{sec:exc}

The goal of this paper is to provide an accurate description of the spectrum of the Heisenberg-Ising ladder Hamiltonian \eqref{eq:ladder} in the 
regime with confining quasiparticles. 
To set clearly the problem, we report the entire spectrum of the Hamiltonian in the sector of zero magnetization for $L=10$,  calculated numerically 
by means of exact diagonalization.
We consider $\Delta_{||}=10$ and $\Delta_\perp=0.1$. 
The spectrum is organized in bands of fixed even number of particles around the energies equal to this number (in units of $\Delta_{||}$). 
In the figure, together with the spectrum of the ladder, we report the numerically calculated spectrum of the XXZ spin-chain in a staggered field which is a mean 
field description of the ladder. As can be seen clearly,
the ladder has many more states than the corresponding chain, which is obvious 
as the Hilbert space of the ladder is exponentially larger than that of the chain.
In the bottom panels, we report zooms of the two-particle and four-particle sectors. 
Inside each band, there is a fine structure given by states with precise quantum numbers. 
Here, we are after an accurate characterization of this fine structure and of the effects of confinement.
A first observation that will have a very simple explanation later is that the two-particle spectrum of the ladder is in one-to-one correspondence
with that of the chain.
This is true modulo a four-fold degeneracy that is a consequence of the discrete symmetries of the Hamiltonian \eqref{eq:HI} (spin flip and chain swap) and is lifted in higher perturbative order in $\epsilon=1/\Delta_{||}$. 
We then notice that the correspondence between the chain and the ladder is not valid for the four-particle sector, where there are many more states that we will describe in the following. 
We stress that the correspondence between the two-particle sectors in the ladder and in the chain does not hold for odd $L.$

In order to understand the structure of the elementary excitations, it is instructive to focus first on the parameter regime $\Delta_{||}\gg 1$.
In this Ising limit it is useful to rescale the Hamiltonian~\eqref{eq:ladder} by $\Delta_{||}$ \cite{Rutkevich2018}, i.e.
\begin{equation}
\label{eq:HI}
	   H_\text{I}(\epsilon, \Delta_\perp) = 
	   \sum_{\alpha=1,2} \sum_{j=1}^{L} \left[\epsilon (\sigma^+_{j,\alpha} \sigma^{-}_{j+1,\alpha} +
	    \sigma^{-}_{j,\alpha} \sigma^{+}_{j+1,\alpha})+\frac{1}{2}(\sigma_{j,\alpha}^{z} \sigma_{j+1,\alpha}^{z} + 1)\right]   
	    + \epsilon\Delta_{\perp}\sum_{j=1}^{L} \sigma^{z}_{j,2} \sigma^{z}_{j,1}\,,
\end{equation}
where $\epsilon=1/\Delta_{||}.$ We study this Hamiltonian perturbatively in $\epsilon.$

\begin{figure}[t]
\centering
\includegraphics[width=0.48\textwidth]{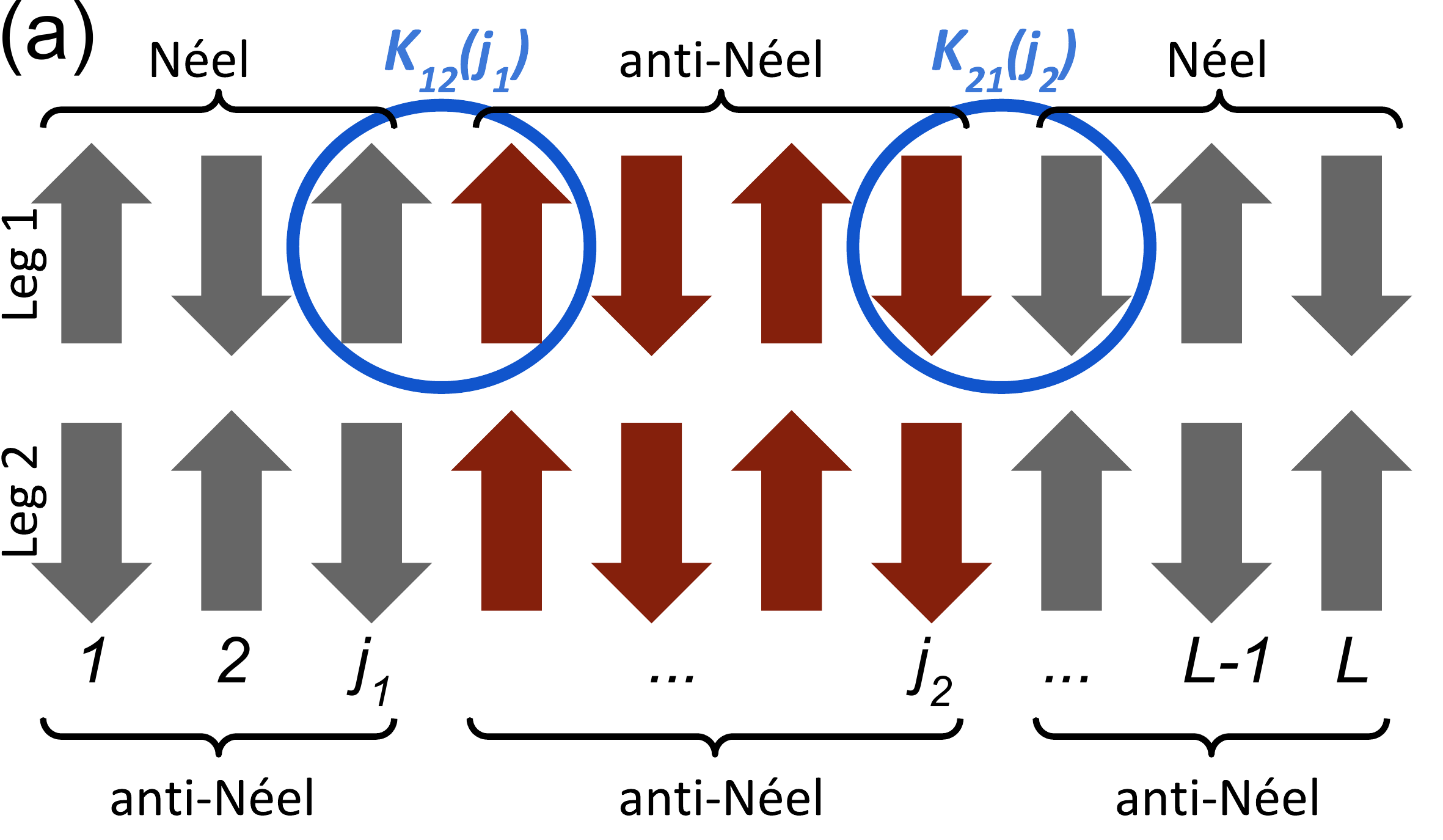}
\hfill
\includegraphics[width=0.48\textwidth]{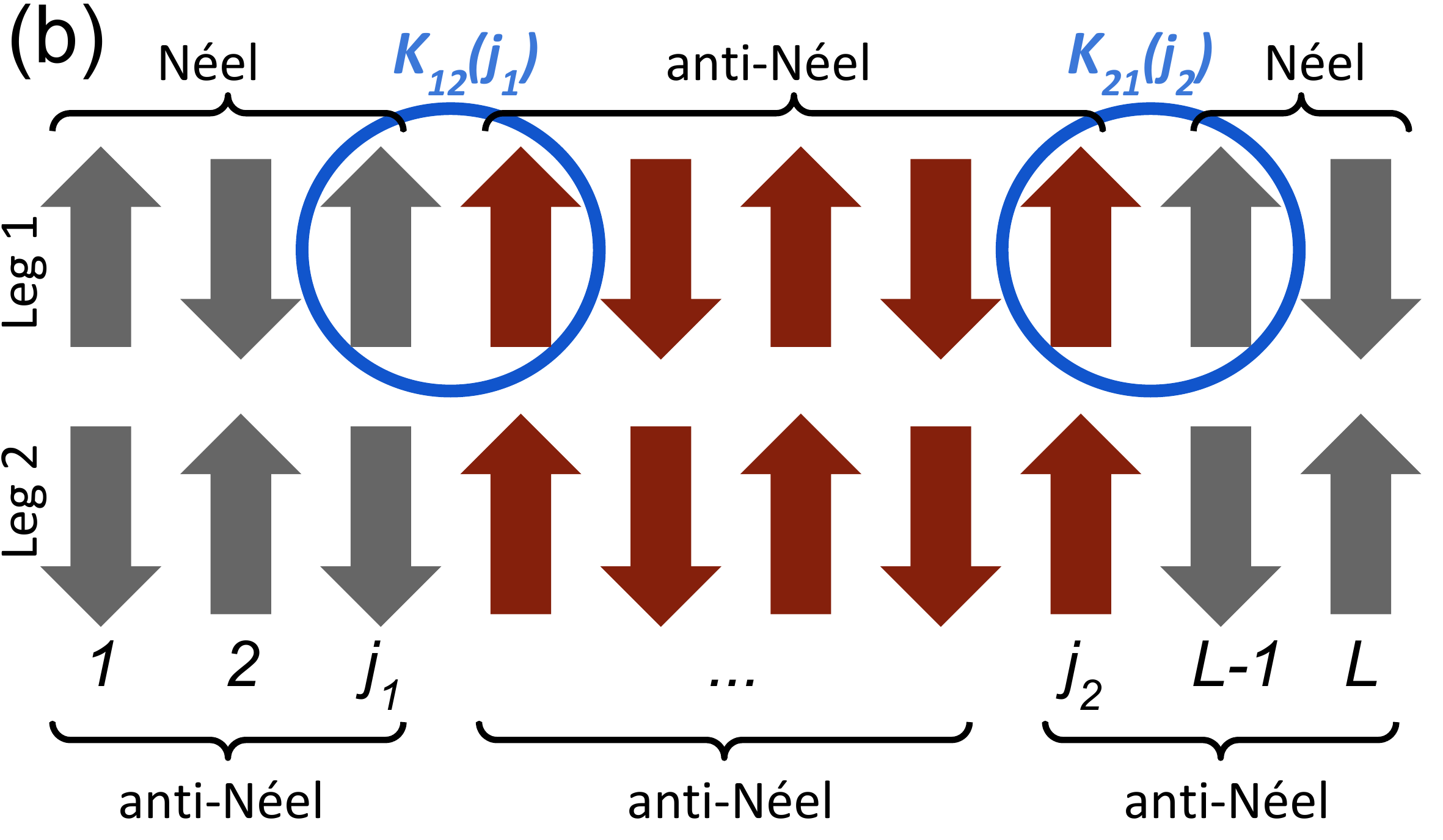}\\
\vspace{4mm}
\includegraphics[width=0.48\textwidth]{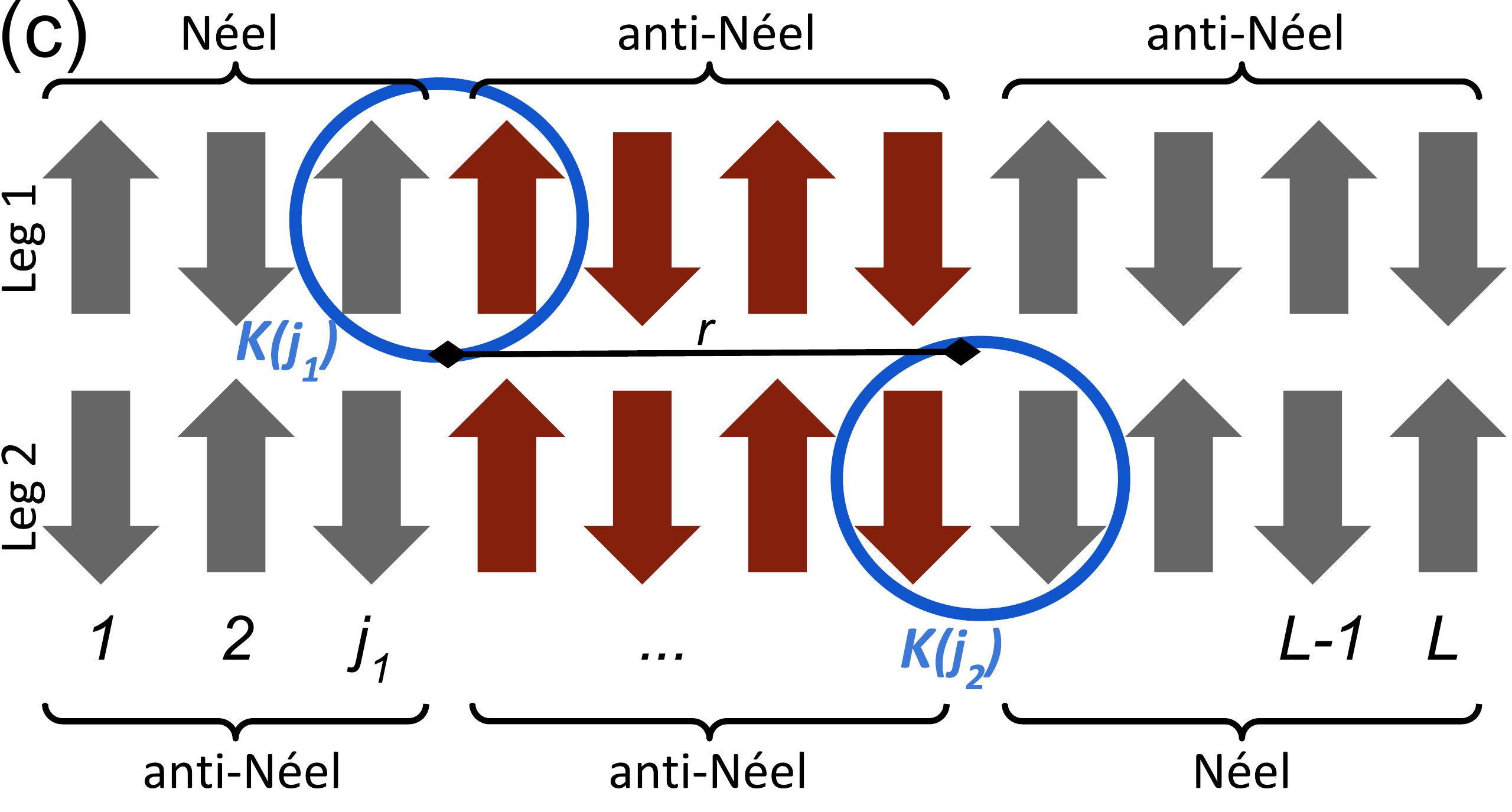}
\hfill
\includegraphics[width=0.48\textwidth]{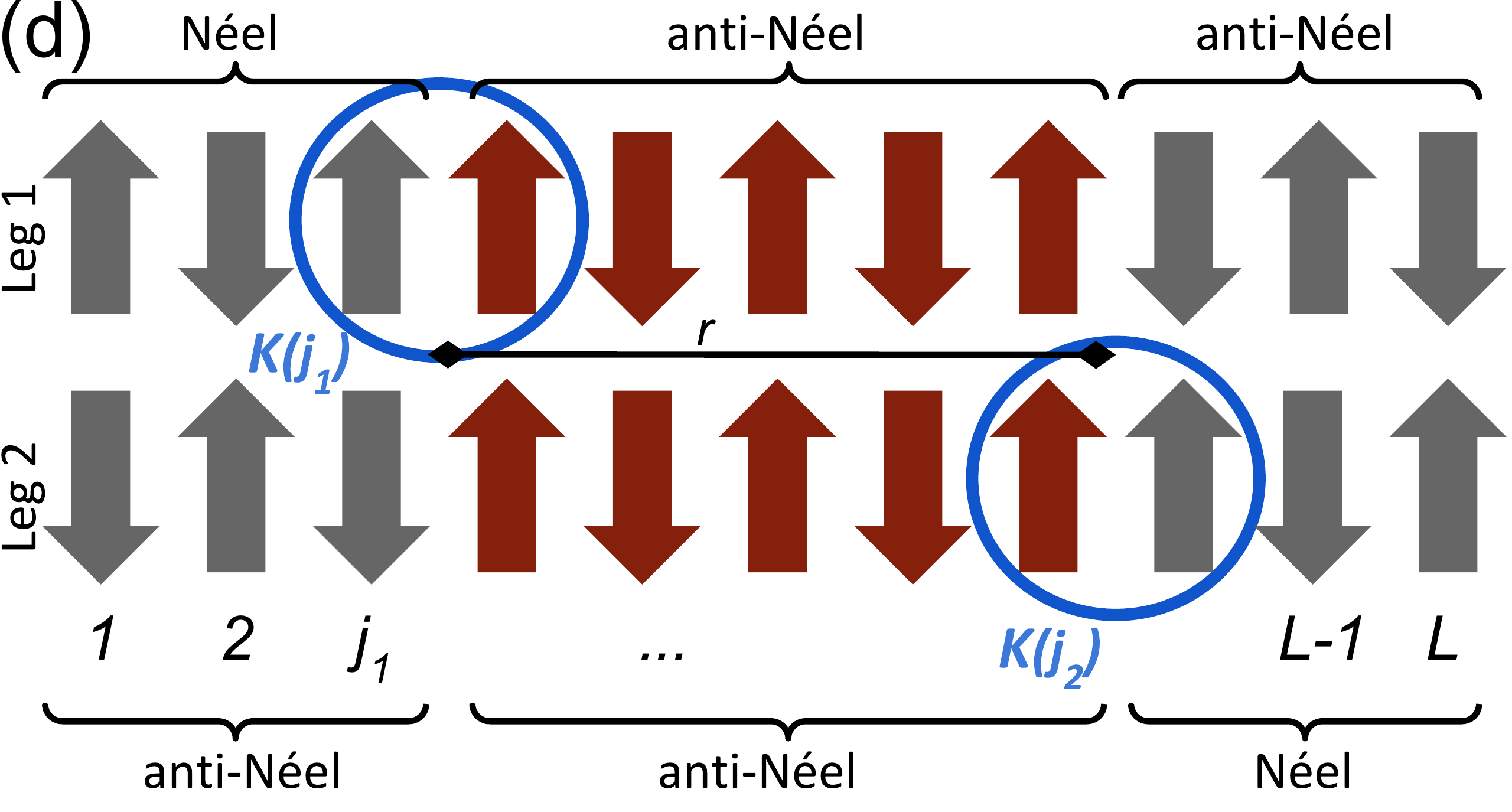}
\caption{Schematic picture of the possible mesons in the ladder. 
In (a) and (b) we have `Type~1' intrachain mesons built from two kinks on the same chain. 
Instead, in (c) and (d) we have `Type~2' interchain mesons built from two kinks on the different chains.  `Type~1' mesons interpolate between the same kind of vacua, while `Type~2' mesons interpolate between vacua of different kind.
The coupling $\Delta_{\perp}$ induces a linear potential between the kinks, 
because the energy cost scales with the distance of the kinks equal to the number of spins that have frustrated interchain links (shown in red). This distance is even for kinks of opposite spins (a,c) and odd for kinks of the same spin (b,d).
}
\label{fig:mesons1}
\end{figure}

When $\epsilon=0,$ the two chains are decoupled and the hopping terms are absent. 
The Hamiltonian has four degenerate ground states given by the four possible combinations of the N\'eel $|\Psi_{1}\rangle = |\uparrow\downarrow\uparrow\downarrow\dots\rangle$ and anti-N\'eel $|\Psi_{2}\rangle = |\downarrow\uparrow\downarrow\uparrow\dots\rangle$ 
states  of the two chains (here $|\!\! \uparrow\rangle$ is chosen with quantization axis in the $z$ direction, i.e., $\sigma^z_j|\!\!  \uparrow\rangle=| \!\! \uparrow\rangle$). 
In the units of Eq. \eqref{eq:HI} all these ground states have exactly zero energy for $\epsilon=0$. 
The fundamental excitations of each chain are kinks $|K_{\alpha\beta}(j)\rangle$
interpolating between the two vacua $\ket{\Psi_\alpha}$ and $\ket{\Psi_\beta}$ ($\alpha,\beta\in\{1,2\},\alpha\neq\beta$) at the bond between sites $j$ and $j+1$ (cf. Fig. \ref{fig:mesons1}). Note that the spin $s$ and the parity of the site $\rho=j\,\mathrm{mod}\,2$ on the chain $\alpha\in\{1,2\}$ are related by $s=(-1)^{\alpha}(1/2-\rho).$
For 
$0<\epsilon\ll1$ and $\Delta_{\perp}=0,$ the exactly known ground states of the chains are still almost N\'eel and anti-N\'eel states,
but their degeneracy is lifted for finite $L$ yielding exponentially small (in $L$) splittings. 
Similarly, the hopping term hybridizes the kink states and lifts their extensive degeneracy. 
When $\Delta_{\perp}>0,$
two of the four 0-kink states in which the spins are anti-aligned along the rungs, gain extensive ($\propto - \Delta_\perp L$) negative energy while the other two gain extensive positive energy ($\propto \Delta_\perp L$). In the thermodynamic limit, the latter two become false vacua and together with all the formerly low lying excitations 
above them are pushed to the top of the spectrum.
On the ladder a single kink in one of the two chains toggles between a true and a false vacuum (illustrated in grey and red, respectively, in Fig.~\ref{fig:mesons1}). 
This implies that the nature of the low-lying excitations change qualitatively when the coupling between the chains is turned on.
The low energy sector only bears states with an even number of kinks since there must be a true vacuum both on the left and on the right of these states. 
Consequently, states made with two kinks become the elementary excitations in the spectrum. 
The energy acquired by the false vacuum between the two kinks induces an effective linear potential between them: 
the two kinks are confined in excitations that we call {\it mesons}, following a standard terminology in the literature.
Because of the presence of two true vacua, we can distinguish between two classes of mesons. 'Type 1' mesons are interpolating between the same vacuum while 'Type 2' mesons are interpolating between two different vacua 
(see Fig. \ref{fig:mesons1} to grasp the idea with a graphical representation).

We now illustrate more clearly the difference between these two mesons in terms of the symmetries of the model.
For $\Delta_{\perp}=0$ the model has a $\mathbb{Z}_2 \times \mathbb{Z}_2$ symmetry associated with the total spin flip along each chain. 
A kink on a given chain has a non-zero $\mathbb{Z}_2$ charge for the spin-flip symmetry on that chain. 
When $\Delta_{\perp}$ is turned on, the symmetry is explicitly broken and only one $\mathbb{Z}_2$ symmetry is left (i.e. the global spin-flip of both chains). 
A charge $Q$ can be assigned to the explicitly broken symmetry: this charge corresponds to the parity of the total number of kinks. 
As a consequence of confinement, the low-energy spectrum can harbour only neutral objects, while charged objects are pushed up in the spectrum. 
Both `Type 1' and `Type 2' mesons are neutral with respect to this symmetry, i.e. they have $Q=0$. 
The remaining $\mathbb{Z}_2$ symmetry is spontaneously broken in the ground state. 
Another charge $q$ may be assigned to this different symmetry: 
`Type 1' mesons correspond to $Q=0$ and $q=0$ while `Type 2' to $Q=0$ and $q=1$. 
Very importantly, since this second symmetry is not explicitly broken, low-energy states do not need to be neutral with respect to $q$. 
On the contrary, it is possible to have charged excitations (`Type 2' mesons) which are the sort of composite kinks for the spontaneously broken global spin-flip symmetry.
While at first, this phenomenon can sound rather peculiar, it is actually very similar to what happens for strong interactions: 
the mesons are neutral particles for the color charge but they are not for the electrical charge, related to another symmetry of nature.

In finite volume $L$, because of periodic boundary conditions, chains with an odd number of sites can host an odd number of kinks while chains with an even number of sites can host an even number of kinks. Namely, for $L$ odd there are only $q=1$ states while for $L$ even only $q=0$ states (the opposite holds for anti-periodic boundary conditions).
Consequently, as long as $\Delta_{\perp}\ll \Delta_{||}$, the lowest energy states are `Type 1' mesons if $L$ is even and `Type 2' mesons if $L$ is odd.
In the following, we give a quantitative account of their dispersion relation both for an infinite system and for a ladder of finite size.
The approach we exploit here is rather standard: we project the many-body Hilbert space onto the 2-kink sector yielding an effective two-body Hamiltonian
which can be treated with elementary quantum mechanics techniques.
As discussed above, the degeneracy of the ground and excited states gets lifted at the first order in $\epsilon$, thus the dispersion relation of the low-energy meson excitations,
which correspond to the low lying many-body levels, can be well described by a first order perturbative analysis in $\epsilon$ restricted to the two-kink sector.

\subsection{`Type 1' intrachain mesons}
\label{sec:Bessel1}
`Type 1' or intrachain mesons are formed by kinks on the same chain, as shown in Fig. \ref{fig:mesons1}.  
In the regime $\Delta_{\perp}\ll \Delta_{||}=\epsilon^{-1},$ the interchain interaction can be studied in a mean field fashion \cite{Bhaseen2004,Schulz1996,Essler1997,Sandvik}, 
by focusing on one of the chains and treating the spontaneous staggered magnetization $\bar{\sigma}$ of the other chain as an effective external field:
\begin{equation}
\label{eq:HS}
	    \hat H_\text{S}(\epsilon,h)= \sum_{j=1}^{L} 
	   \left[  \epsilon\left(\sigma^+_j \sigma^-_{j+1} +\sigma^-_j \sigma^+_{j+1}\right)
	    + \frac{1}{2}(\sigma_j^z \sigma_{j+1}^z + 1)\right]
	    + \epsilon h \sum_{j=1}^{L} (-1)^j \sigma_j^z\,.
\end{equation}
where $h=\bar{\sigma} \Delta_{\perp}$. 
Here we assume that the other chain is in the approximate anti-N\'eel state; the N\'eel case follows by the global spin flip symmetry.
In the limit $\Delta_{||}\gg 1,$ 
the staggered magnetization is $\bar\sigma\approx 1$.
The excitations of the infinite antiferromagnetic chain (i.e. Hamiltonian \eqref{eq:HS} with $L\to\infty$) and their confinement have been studied 
using various approximations in Ref. \cite{Rutkevich2018}. Here we extend the analysis of this work to finite chains.

We introduce the projector $\hat P_2$ onto the 2-kink subspace spanned by the basis $|K_{\alpha\beta}(j_1)K_{\beta\alpha}(j_2)\rangle$. 
The action of the projected Hamiltonian $\hat H_2=\hat P_2 \hat H_{\text{S}} \hat P_2$ on 2-kink states is easily worked out as
\begin{equation}
\begin{split}
\hat H_2(\epsilon, \Delta_{\perp})&|K_{\alpha\beta}(j_1)K_{\beta\alpha}(j_2)\rangle =\\
&\left[2+(-1)^\alpha \epsilon h(L-2 j)\right] |K_{\alpha\beta}(j_1)K_{\beta\alpha}(j_2)\rangle\\
+\epsilon \big\{&\left[|K_{\alpha\beta}(j_1-2)K_{\beta\alpha}(j_2)\rangle+|K_{\alpha\beta}(j_1)K_{\beta\alpha}(j_2+2)\rangle\right](1-\delta_{ j,L-1})(1-\delta_{j,L-2})\\
+&\left[|K_{\alpha\beta}(j_1)K_{\beta\alpha}(j_2-2)\rangle+|K_{\alpha\beta}(j_1+2)K_{\beta\alpha}(j_2)\rangle\right](1-\delta_{ j,1})(1-\delta_{ j,2})\big\}\,,
\end{split}
\end{equation}
 where $1\le j=j_2-j_1\le L-1$. The first line gives the effective potential, while the second and third lines describe the hopping of the kinks by two sites. The Kronecker-delta factors encode the hard-core nature of the kinks.

Exploiting translational invariance by 2 sites, we are looking for the energy eigenfunctions in the sector of total spin $s$ in the form
\begin{subequations}
\begin{align}
|\Psi_{n}(s=\pm1)\rangle &= \sum_{j_1=\frac{3-s}2}^{L-\frac{1+s}2} {}^{'} \sum_{r=1}^{L-1}{}^{'}\psi^{(s)}_n(r|P)e^{iP(j_1+r/2)}| K_{12}(j_1)K_{21}(j_1+r)\rangle\,,\\
|\Psi_{n}^\text{(o)}(s=0)\rangle &= 
\sum_{j_1=1}^{L-1}{}^{'} \sum_{r=2}^{L-2}{}^{'}\psi^{\text{(o)}}_n(r|P)e^{iP(j_1+r/2)}| K_{12}(j_1)K_{21}(j_1+r)\rangle\,,\\
|\Psi_{n}^\text{(e)}(s=0)\rangle &=
\sum_{j_1=2}^{L}{}^{'} \sum_{r=2}^{L-2}{}^{'}\psi^{\text{(e)}}_n(r|P)e^{iP(j_1+r/2)}| K_{12}(j_1)K_{21}(j_1+r)\rangle\,,
\end{align}
\end{subequations}
where the primed sums run over odd or even integers, and the momentum of the center of mass  $P$ is quantized as $P=k\,2\pi/L$, 
$k=-\left \lfloor{\frac{L}4}\right \rfloor,\dots, \left \lfloor{\frac{L}4}\right \rfloor.$ The limited range of the momentum reflects the doubling of the unit cell due to the staggered background field. The ``o/e'' superscripts refer to the odd and even sites on which the kinks are located. The parity of the distance $r$ is fixed by the spin of the kinks: $r=2,4,\dots L-2$ if the total spin is 0 and $r=1,3,\dots L-1$ if the total spin is $\pm 1$ (see Fig.~\ref{fig:mesons1}).

\begin{figure}[!t]
\centering
\includegraphics[width=1\textwidth]{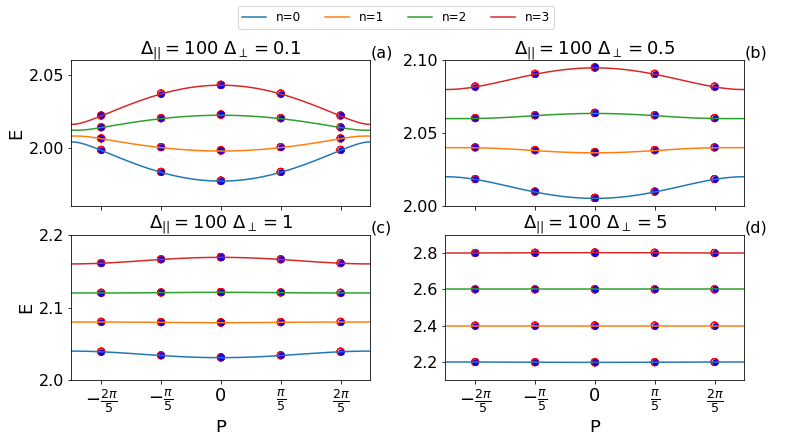}
\caption{  Low-lying part of the spectrum in the spin $s=0$ sector of the ladder in Eq.~\erf{eq:HI} (red circles) and of the staggered XXZ chain in Eq.~\erf{eq:HS} (blue dots) for $L=10,$ $\Delta_{||}=100$, and $\Delta_\perp=0.1,0.5,1,5.$  
The numerical data have been obtained by exact diagonalization.
The dispersion relations of the mesons in the 2-kink approximation \erf{eq:En1} are shown in continuous lines, obtained by solving Eq.~\erf{eq:type1s0} numerically. 
The internal quantum number $n$ of each curve is in the legend on top of the plot. 
In the strong anisotropy regime, and for even $L$, the ladder is equivalent to the staggered XXZ chain in the two-kink sector.
 }
\label{fig:type1delta100}
\end{figure}

Using these expressions, the eigenvalue problem of $\hat H_{2}$ leads to the discrete Sturm--Liouville equation
 \begin{equation}
 \label{eq:type1s0}
(2+2\epsilon \Delta_{\perp}\,r)\psi^{(a)}_n(r) + 2 \epsilon \cos(P)\big[ \psi^{(a)}_n(r+2)+\psi^{(a)}_n(r-2)\big]=E^{(a)}_n(P)\,\psi^{(a)}_n(r),
 \end{equation}
for all relative wave functions $\psi^{(a)}_n(r|P),$ $a\in\{+1,-1,\text{o},\text{e}\}.$ Here $E_n$ are the excitation energies with respect to the ground state energy $\mathcal{E}_\text{GS}=-\epsilon \Delta_{\perp} L.$
The boundary conditions are $\psi^{(\pm1)}_n(-1)=\psi^{(\pm1)}_n(L+1)=0$ and $\psi^{(\text{o/e})}_n(0)=\psi^{(\text{o/e})}_n(L)=0.$ 
The solutions can be written down exploiting the recurrence relation satisfied by the Bessel functions of the first $J_{\nu+1}(z)+J_{\nu-1}(z) = 2\nu/z\,J_{\nu}(z)$ and 
similarly for and the second kind $Y_{\nu}(z)$, obtaining
\begin{equation}
\psi^{(a)}_n(r|P) = N^{(a)}_n\left[ J_{\nu^{(a)}_n(P)-r/2}(\Delta_{\perp}^{-1}\cos P) + A^{(a)}_n\,Y_{\nu^{(a)}_n(P)-r/2}(\Delta_{\perp}^{-1}\cos P) \right]\,,
\end{equation}
where $N^{(a)}_n(P)$ is the normalization and $\nu^{(a)}_n(P)$ and $A^{(a)}_n(P)$ are determined by the boundary conditions. 
These solutions are labeled by the integer $n$ and their energy eigenvalues are 
\begin{equation}
\label{eq:En1}
E^{(a)}_n(P) = 2 + 4 \epsilon \Delta_{\perp} \,\nu^{(a)}_n(P)\,.
\end{equation}
We plot the energy levels $E_n(P)$ obtained by solving Eq.~\erf{eq:type1s0} (or equivalently  
Eq. \erf{eq:En1}) for $L=10$ in the case of total spin $s=0$ in Fig.~\ref{fig:type1delta100}.
These analytic predictions are compared to the exact diagonalization results both for the ladder Hamiltonian \erf{eq:HI} and for the staggered chain 
Hamiltonian \erf{eq:HS} for $\Delta_{||}=100$ and different values of the interchain coupling $\Delta_\perp$. 
For these couplings, the low energy part of the ladder spectrum matches perfectly the spectrum of the staggered XXZ chain. Moreover, both spectra are well captured by the 2-kink approximation.

\begin{figure}[!t]
\centering
\includegraphics[width=1\textwidth]{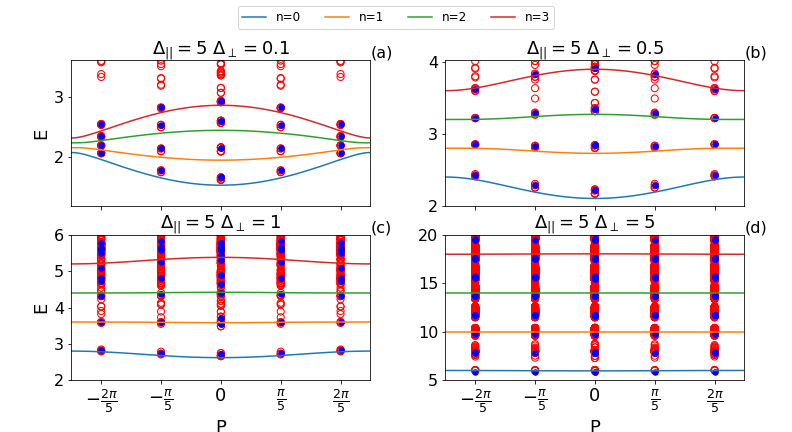}
\caption{ Low-lying part of the spectrum in the spin $s=0$ sector of the ladder in Eq.~\erf{eq:HI} (red circles) and of the staggered XXZ chain in Eq.~\erf{eq:HS} (blue dots) obtained by exact diagonalization for $L=10$, $\Delta_{||}=5,$ and $\Delta_\perp=0.1,0.5,1,5.$  The dispersion relations of the mesons in the 2-kink approximation are shown in continuous lines, obtained by solving Eq.~\erf{eq:type1s0} numerically. The internal quantum number $n$ is in the legend on top of the plot.}
\label{fig:type1delta5}
\end{figure}

In Fig.~\ref{fig:type1delta5} we explore the robustness of this effective description as we moveaway from the strong anisotropic region
by reporting a comparison with the numerical results from exact diagonalization for a ladder of length $L=10$ with $\Delta_{||}=5.$
Even though the quantitative agreement is worse for $\Delta_{||}=5$ than for $\Delta_{||}=100,$
the effective two-kink Hamiltonian still represents a good qualitative description of the low energy states as long as the energy bands of different kink numbers are well separated.
Indeed, the main qualitative effect is that as $\Delta_\perp$ increases (at fixed $\Delta_{||}$), some high energy states, which are not captured 
by the mean-field staggered XXZ chain, come down to low energy and mix up (and at some point hybridize) with the part of the spectrum we are able to describe.
At a more quantitative level, even for the smallest values of $\Delta_\perp$ we observe deviations that anyhow were expected.
Indeed, as $\Delta_{||}$ is decreased, the fundamental excitations interpolate between vacua that cannot be approximated by a N\'eel or an anti-N\'eel state. 
Moreover, the nontrivial scattering properties of those excitations will start to play a role. 
Both effects will be investigated in the next section.

\subsection{`Type 2' interchain mesons}
\label{sec:mesontype2}

We now turn to the meson excitations that are formed by two kinks located on different chains (see Fig.~\ref{fig:mesons1}). On a ladder with periodic boundary condition, these states can only exist for $L$ odd, and they have no equivalent in a staggered XXZ chain.

We can follow steps very similar to those for intrachain mesons in the previous subsection. 
We first project onto states having one kink on each leg of the ladder.
The main difference is that in this case there is no hard-core constraint for the kinks as they can cross by passing above/below each other and hence their 
wave function is
\begin{equation}
|\Psi_n(s)\rangle = \sum_{j_1=1}^L \sum_{r=0}^{L-1}\phi_n(r|P,s)e^{iP(j_1+r/2)}| K(j_1)\rangle_1 |K(j_1+r)\rangle_2\,,
\end{equation}
where the subscripts $1,2$ label the legs of the ladder.  The spins of the kinks should add up to $s$. 
The center of mass momentum $P$ is quantized as $P=k\,2\pi/L,$ $k=-\left \lfloor{\frac{L}2}\right \rfloor,\dots, \left \lfloor{\frac{L}2}\right \rfloor$.

\begin{figure}[!t]
\centering
\includegraphics[width=1\textwidth]{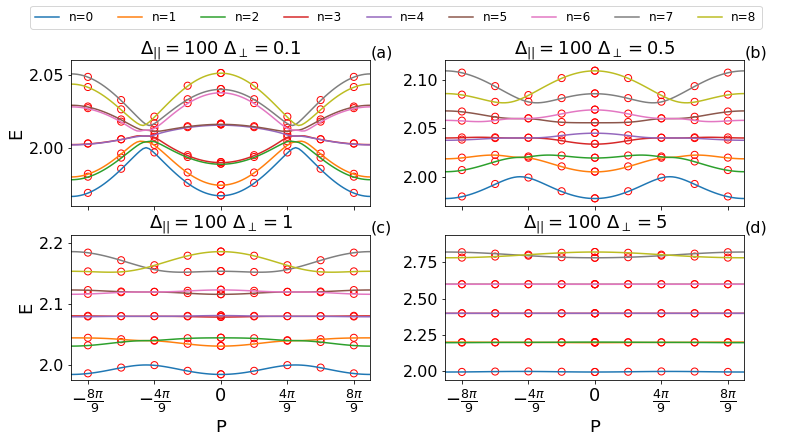}
\caption{  Low-lying part of the spectrum in the spin sector $s=0$ of the ladder \erf{eq:HI} (red circles) for $L=9,$ $\Delta_{||}=100,$ and $\Delta_\perp=0.1,0.5,1,5.$  The dispersion relations of the mesons in the 2-kink approximation \erf{eq:En2} are shown in continuous lines, obtained by solving Eq. \erf{eq:type2} numerically.
The internal quantum number $n$ is in the legend on top of the plot.
Notice many qualitative different features compared to the ladder of even length in Fig. \ref{fig:type1delta100}.}
\label{fig:type2delta100}
\end{figure}

 The equation for the relative wave function in all spin sectors turns out to be
\begin{equation}
\label{eq:type2}
     (2+2\epsilon \Delta_{ \perp}\ell_s(r))\phi_n(r) + 2 \epsilon \cos(P)\left[ \phi_n(r+2)+\phi_n(r-2)\right]=E_n(P)\,\phi_n(r),
 \end{equation}
 with $r=0,1,\dots L-1$.
 The function $\ell_s(r)$ is the length of the string between the kinks and is defined as
 \begin{equation}
 \label{eq:ell}
     \ell_{s=0}(r)=\begin{cases}
        r \text{ if $r$ even}\\
        L-r \text{ if $r$ odd,}
     \end{cases}
     \qquad\qquad
      \ell_{s=\pm 1}(r)=\begin{cases}
        r \text{ if $r$ odd}\\
        L-r \text{ if $r$ even.}
     \end{cases}
 \end{equation}
Note that swapping the chains corresponds to $r\leftrightarrow L-r$ which also changes the parity of $r$,
thus the definitions \eqref{eq:ell} are consistent with this symmetry.
The boundary conditions are $\phi_n(-2)=\phi_n(L-2),$ $\phi_n(2)=\phi_n(L+2),$ and $\phi_n(-1)=\phi_n(L-1),$ $\phi_n(1)=\phi_n(L+1).$  The solutions are given by 
\begin{multline}
\phi^{(\pm)}_n(r|P,s) = (\pm)^r N^{(\pm)}_n(P,s)\left[ J_{\nu^{(\pm)}_n\!(P,s)-\ell(r)/2}(\Delta_\perp^{-1}\cos P) +
\right. \\+ \left.
A_n^{(\pm)}(P,s)\,Y_{\nu^{(\pm)}_n\!(P,s)-\ell(r)/2}(\Delta_\perp^{-1}\cos P) \right]\,,
\end{multline}
where $N^{(\pm)}_n(P,s)$ is the normalization, $\nu^{(\pm)}_n(P,s)$ and $A^{(\pm)}_n(P,s)$ are fixed by the boundary conditions which now relate the wave function at odd and even sites. 
Notice that the four boundary conditions give only two independent equations. 
$\phi^{(+)}_n(r)$ are symmetric, while $\phi^{(-)}_n(r)$ are anti-symmetric under the exchange $r\leftrightarrow L-r.$
The energies are given by
\begin{equation}
\label{eq:En2}
E^{(\pm)}_n(P,s) = 2 + 4 \epsilon \Delta_\perp \,\nu^{(\pm)}_n(P,s)\,.
\end{equation}

\begin{figure}[!t]
\centering
\includegraphics[width=1\textwidth]{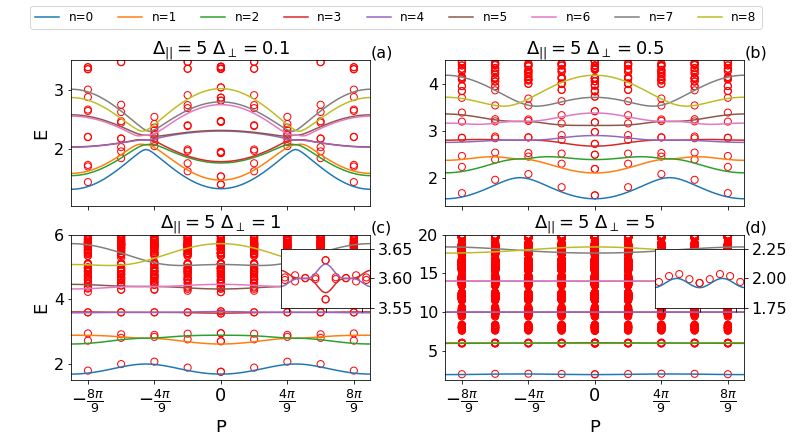}
\caption{  Low-lying part of the spectrum in the spin sector $s=0$ of the ladder \erf{eq:HI} (red circles) for $L=9,$ $\Delta_{||}=5,$ and $\Delta_\perp=0.1,0.5,1,5.$  The dispersion relations of the mesons in the 2-kink approximation \erf{eq:En2} are shown as continuous lines, obtained by solving Eq.~\erf{eq:type2} numerically. 
The internal quantum number $n$ is in the legend on top of the plot.
The insets show the accuracy of our approximation in resolving the spectrum on a more refined scale.}
\label{fig:type2delta5}
\end{figure}

In Fig.~\ref{fig:type2delta100} we compare the levels obtained from Eq.~\erf{eq:En2} with the results of exact diagonalization of the ladder Hamiltonian for $L=9$ in the spin sector $s=0$ for $\Delta_{||}=100$ and different values of the interchain coupling $\Delta_\perp.$ Similarly to the intrachain mesons, the spectrum is very well captured by the effective 2-kink description. The figure also demonstrates the richer structure of the `Type 2' interchain mesons as they have about twice as many internal excitations as the `Type 1' intrachain mesons have.

In Fig.~\ref{fig:type2delta5} the same comparison is shown for $\Delta_{||}=5$ and various values of $\Delta_{\perp}$. Analogously to the case of `Type 1' mesons, the overall structure of the spectrum is captured by the 2-kink approximation in the regions where the bands are well separated. 
The deviations of exact numerical results from the 2-kink approximation are more pronounced than for $\Delta_{||}=100$, because the dressing of the fundamental excitations becomes relevant for small $\Delta_{||}$.
    
We conclude this section by mentioning that bound states between two coupled 1+1 dimensional models have been observed also in conformal field theories
\cite{muss98}, but in a very different context that does not lead to confinement.

\section{A semiclassical approach for finite $\Delta_{||}$}
\label{sec:BS}

When approaching smaller values of the anisotropy parameter $\Delta_{||}$ at $\Delta_{\perp}=0,$ the ground state of the model and its fundamental excitations experience significant changes. The doubly degenerate ground states of both chains still have anti-ferromagnetic order but with a smaller average staggered magnetization. The latter is exactly known from the Bethe ansatz solution of the XXZ spin chain and it is given by 
\begin{equation}
\label{eq:stagmag}
\bar{\sigma} = \prod_{n=1}^{\infty} \left( \frac{1-e^{-2n\gamma}}{1+e^{-2n\gamma}} \right) ^{2}\,, 
\qquad   \Delta_{||}=\cosh(\gamma )\,.
\end{equation}
As $ \Delta_{||}$ decreases, the ground states with the above staggered magnetization are no longer well approximated by a N\'eel or an anti-N\'eel state. 
The elementary excitations are still topological quasi-particles that interpolate between the two vacua, but now they interact in a nontrivial way. In this section we describe how these properties affect the `Type 1' intrachain mesons.

The most pragmatic way to treat the presence of a non-vanishing $\Delta_\perp$ would be a perturbative expansion in small $\Delta_{\perp}$  around the exact eigenstates at $\Delta_{\perp}=0$. The latter approach is rather technical and involves a Bethe--Salpeter equation with a perturbative form factor expansion. 
Although less rigorous, here we follow another, more heuristic approach whose main advantage is having a straightforward physical interpretation. 
Following Ref. \cite{Rutkevich2008}, the idea is to
look for semi-classical bound states of the Hamiltonian
\begin{equation}
\label{eq:classham}
H(x_1,x_2,\th_1,\th_2) = \omega(\th_1)+\omega(\th_2)+f(|x_2-x_1|)\,,
\end{equation}
where 
\begin{equation}
f=2\epsilon \Delta_\perp \bar\sigma^2,
\end{equation}
is the ``string tension'' taking into account the average magnetization of both chains,
and we introduced the continuum coordinates $x_1,x_2 \in \mathbb{R}$ and their canonical conjugate momenta $\th_1, \th_2$. The function $\omega(\th)$ is the lattice dispersion relation of the kink quasi-particle obtained by a Bethe Ansatz approach \cite{zabrodin1992}
\begin{equation}
\label{eq:disp}
    \omega(\th)= \frac{2 \epsilon  K(k)}{\pi} \sinh{\gamma \sqrt{1-k^2 \cos^2 \th}}\,,
\end{equation}
where $K(k)$ is the complete elliptic integral whose modulus $k$ is related to the anisotropy through the relation $K(\sqrt{1-k^2})/K(k)=\gamma /\pi.$ The Hamiltonian \eqref{eq:classham} describes the classical motion of two particles experiencing a long range interaction with kinetic energy given by the exact kinetic energy of a kink of the XXZ chain.
After a canonical transformation to center of mass and relative coordinates, 
\begin{align}
X&=\frac{x_1+x_2}2\,,& x&=x_2-x_1\,,\\
\Th&=\th_1+\th_2\,,& \th&=\frac{\th_2-\th_1}2\,,
\end{align}
the Hamiltonian reads
\begin{equation}
H(X,\Th;x,\th) = \eps(\th|\Th)+f|x|,
\end{equation}
with $
\eps(\th|\Th) = \omega(\Th/2-\th) + \omega(\Th/2+\th)
$.
In these new variables, the equations of motion are
\begin{align}
\dot X &= \frac{\partial\eps}{\partial \Th}\,,& \dot \Th &= 0\,,\\
\dot x &= \frac{\partial\eps}{\partial \th}\,,& \dot \th &= -f \,\mathrm{sgn}(x)\,.
\label{eq:canoniceq}
\end{align}
The total momentum $\Th$ is an integral of motion as well as the energy
\begin{equation}
\label{eq:energycons}
E = \eps(\th|\Th)+f|x|\,.
\end{equation}

\begin{figure}[t]
\centering
\includegraphics[width=.6\textwidth]{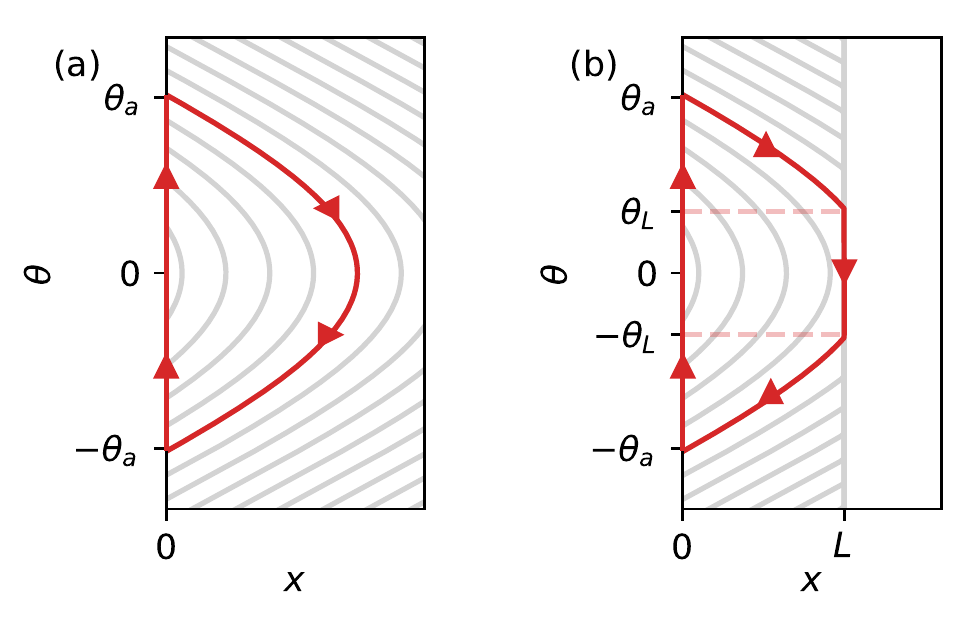}
\caption{Classical trajectories in phase space $(x,\theta)$ for the Bohr--Sommerfeld quantization. The grey lines are the various trajectories at fixed energy.  
In infinite volume, the trajectories are always like the red one depicted in (a). 
In finite volume $L$ we have a hard cutoff that deforms some trajectories as reported in (b).  
}
\label{fig:tra}
\end{figure}

The bound state energies can be obtained via the Bohr--Sommerfeld quantization condition which reads as
\begin{equation}
\label{eq:bohrsomm}
\oint \th dx = 2\pi(n+\delta)\,,\qquad\qquad n=0,1,2,\dots,
\end{equation}
where the integral is taken over the closed path in the $(\th,x)$ classical phase space, and $\delta$ is a phase shift discussed below. In principle, the energies obtained from this equation become more and more accurate with increasing $n$. For simplicity, we restrict the analysis to the case when $\eps(\th|\Th)$ has a single minimum at $\th=0$ which holds for $\Th<\Th_\text{c} =\arccos\left(\frac{1-\sqrt{1-k^2}}{1+\sqrt{1-k^2}}\right)$ \cite{Rutkevich2018}. 
The path is an arc in the $x>0$ half plane (see Fig. \ref{fig:tra}-a) parameterized using Eqs.~\eqref{eq:energycons} and \eqref{eq:canoniceq} starting from $x(0)=0^+ ,\th(0)=\th_a:$
\begin{align}
x(t)&=\frac{E-\eps (\th(t)|\Th)}{f}\,, \\
\th(t)&= \th_a - f t\,,
\end{align}
where $\th_a$ satisfies $E=\eps(\th_a|\Th).$ 
The turning point is at
\begin{equation}
\label{eq:xmax}
x_\text{max}=\frac{E-\eps(0|\Th)}f,
\end{equation}
and is reached at time $t_\text{max}=\th_a/f.$ After another $t_\text{max}$ time elapses, the two kinks scatter at $x(2t_\text{max})=0$ which abruptly flips the sign of $\th,$
so the phase space path is closed by a straight segment at $x=0$ connecting $-\th_a$ with $\th_a.$
These phase space paths are reported in Fig. \ref{fig:tra}-a.
The left hand side of Eq.~\eqref{eq:bohrsomm} reads
\begin{equation}
    \begin{split}
    \oint \th dx & = -\int_{-\th_a}^{\th_a} d\th \th \frac{d x(\th)}{d\th}= -\frac{1}{f} \int_{-\th_a}^{\th_a} d \th \th \dot x(\th) \\
    &= \frac{1}{f} \int_{-\th_a}^{\th_a} d \th \th \frac{\partial \eps (\th|\Th)}{\partial \th}=  \frac{1}{f}\left(2E \th_a  -\int_{-\th_a}^{\th_a} d \th \eps (\th|\Th)\right)\,,
    \end{split}
\end{equation}
where Eq.~\eqref{eq:canoniceq} was used to trade the time derivative for a derivative with respect to $-\th.$

\begin{figure}[!t]
\centering
\includegraphics[width=1\textwidth]{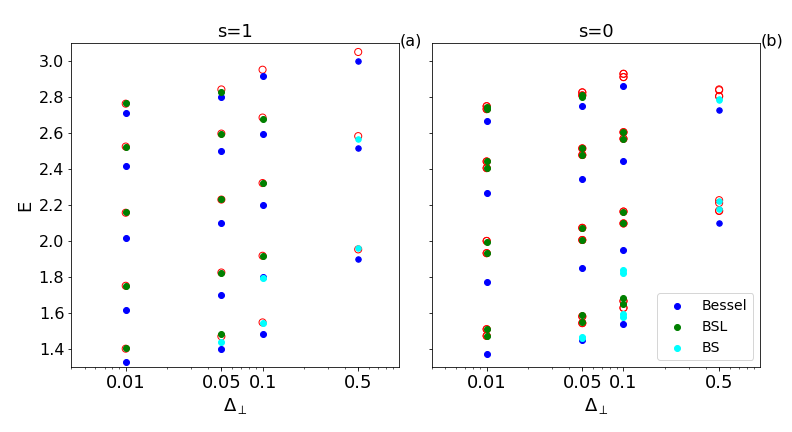}
\caption{Comparison of various approximations for the intrachain meson energies at $\Delta_{||}=5$ in the sectors with spin $s=0$ (right) and $s=1$ (left), both with total momentum $P=0$.  The exact diagonalization results are shown in red empty circles. 
The full symbols correspond to the 2-kink approximation of Sec. \ref{sec:Bessel1} (``Bessel"), the infinite volume semiclassical (``BS") and the finite volume semiclassical (``BSL") results according to the color code shown in the legend.}
\label{fig:BSdelta5}
\end{figure}

The phase shift $\delta$ receives two kinds of contributions. First, at the regular turning point there is a $\pi/2$ phase shift ($\delta=1/4$). Second, at $x=0$ we have to take into account the scattering phase shift of the particles. In the Ising regime $\Delta_{||}\gg1,$ the kinks behave as free hard core particles, so their scattering phase shift is simply $\pi.$ This is equivalent to enforcing that the relative wave function vanish at the origin, and leads to $\delta=1/2$, the same as for a particle suffering a hard reflection. Away from the Ising limit the kinks have a nontrivial, momentum-dependent scattering phase shift $\phi_\eta(p_1,p_2)$ that can be obtained via Bethe ansatz (see Appendix \ref{app:BA} for its detailed expression). The index $\eta$ accounts for the spins of the kinks and will be dropped from now on to simplify the notation. This phase needs to be added to the left hand side of Eq.~\eqref{eq:bohrsomm}, which leads to 

\begin{equation}
\label{eq:BS}
2E(\th_a) - \int_{-\th_a}^{\th_a} dq\,\eps(\th|\Th) = 2\pi f\left(n+\frac34\right) + f \,\phi \left(\frac\Th2-\th_a, \frac\Th2+\th_a\right)\,.
\end{equation}

The procedure has to be modified in finite volume, when the maximum separation $x_\text{max}$ can become larger than the system size $L$. 
There are two possible cases depending on the value of $x_\text{max}.$ If $x_\text{max}<L,$ then the energy levels are given by the solutions of Eq.~\eqref{eq:BS}, while for $x_\text{max}>L$ the system  does not reach the turning point but experiences another scattering at $x=L.$ 
The integration paths in phase-space are shown for the two cases in Fig. \ref{fig:tra}-b.

Let us  denote by $\th_L$ the momentum right before the reflection at $x=L.$ Then the momentum $\th$ jumps from $\th_L$ to $-\th_L$ so the arc in the phase space is chopped to have a flat part at $x=L.$ As a consequence, the new quantization equation reads
\begin{multline}
\label{eq:BSL}
2f\th_L L +2 E(\th_a-\th_L) - 2\int_{\th_L}^{\th_a} dq\,\eps(q|\Th) \\
= 2\pi f\, (n+1) + f \,\phi \left(\frac\Th2-\th_a, \frac\Th2+\th_a\right)+ f \,\phi \left(\frac\Th2-\th_L, \frac\Th2+\th_L\right),
\end{multline}
together with the conditions $\eps(\th_a|\Theta) = E$ and $\eps(\th_L|\Theta) = E-fL.$

We compare the predictions for the intrachain meson energies with spin $s=0$  at $P=\Th=0$ (meson mass gaps) of the 2-kink effective equation \eqref{eq:type1s0} 
and those of the semiclassical quantization \eqref{eq:BS}, \eqref{eq:BSL} with exact diagonalization data of the ladder at $\Delta_{||}=5$ in Fig. \ref{fig:BSdelta5}. 
The plot clearly shows that while the 2-kink approximation with hard-core kinks breaks down, the (finite volume) semiclassical approximation yields an excellent agreement with the exact diagonalization results. Thanks to the spin-dependent phase shift, this approximation also predicts the energy splitting for the case $s=0$, which partially lifts the degeneracy of the spectrum (from four-fold to two-fold) in the thermodynamic limit and cannot be captured with the first approach. Remarkably, even though the semiclassical method is supposed to work well for high energy bound states with large quantum numbers, it gives accurate results even for the lowest lying mesons.

\section{Composite excitations}
\label{sec:compexc}


We recall that in the regime $\Delta_{||}\gg1$ and $\Delta_{||}\gtrsim\Delta_\perp$, the energy spectrum of the Hamiltonian \eqref{eq:ladder} is organized in bands of 
states with a given number of kinks, as shown in Fig. \ref{fig:tot}.
In the previous two sections, we developed an effective systematic description for the low lying 2-kink states. 
Here, we introduce  a more heuristic treatment to grasp the nature of some of the higher excited states. 
We focus on the case of even $L$ and zero magnetization in both chains, i.e. $M_1=M_2=0$.

\begin{figure}[!t]
\centering
\includegraphics[width=0.45\textwidth]{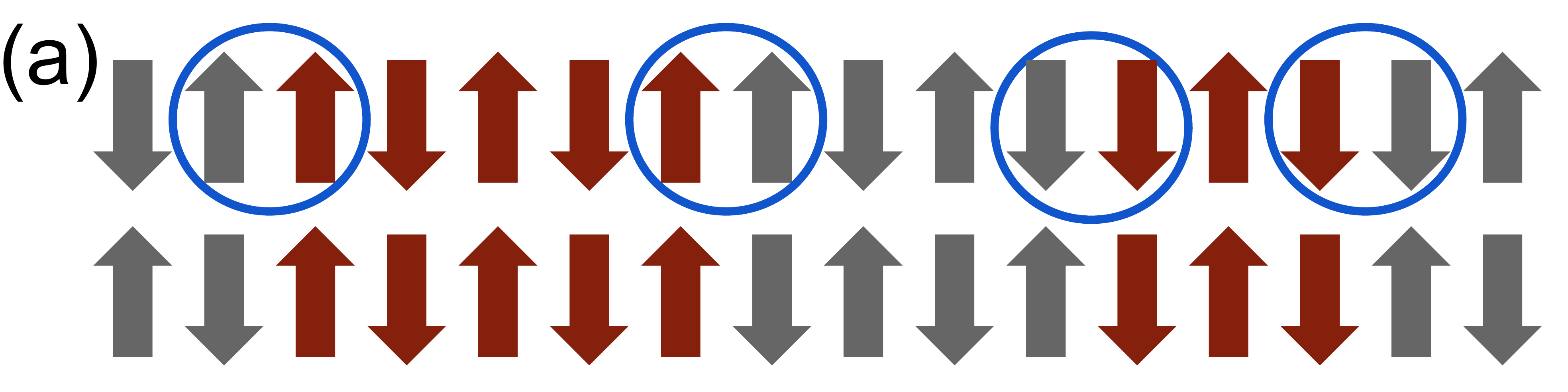}
\hfill
\includegraphics[width=0.45\textwidth]{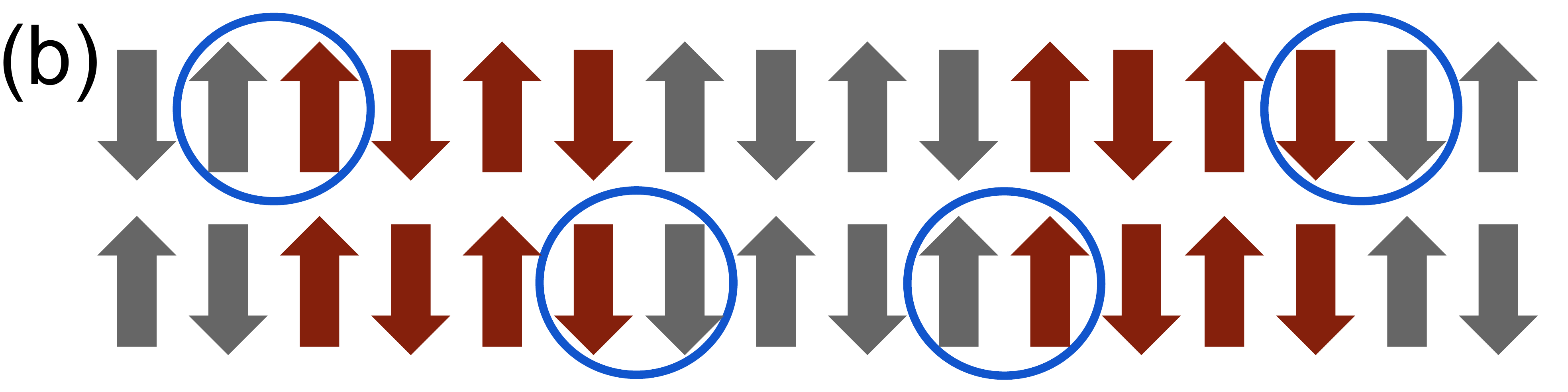}
\caption{Schematic pictures of 4-kink states for a ladder of even length: 
(a) a 4-kink state composed of two intrachain mesons and (b) a 4-kink state composed of two interchain mesons.
}
\label{fig:4kink2}
\end{figure}

The first class of states lying above the 2-kink ones are obviously the four-kink states.
For even $L$, the latter can either be a combination of two intrachain (Fig.~\ref{fig:4kink2}-a) or of two interchain mesons (Fig.~\ref{fig:4kink2}-b).
Notice that for odd $L$ we instead have only combinations of one intrachain and one interchain meson, a situation that we do not describe here.
In the zeroth order approximation in which we neglect the interaction between mesons, the energy levels of the four-kink states is 
just the sum of the dispersion relations obtained in 
the preceding sections for the single mesons, i.e.
\begin{equation}
\label{eq:4kink}
E(n_1,n_2,p,P)= \frac{1}{2} \left[E_{n_1}\left(\frac{P}{2}+p\right)+
E_{n_2}\left(\frac{P}{2}-p\right) + \{n_1\leftrightarrow n_2\} \right]\,,
\end{equation}
where $n_{1}$ and $n_{2}$ collectively label the internal states of the mesons, 
$p=(p_1-p_2)/2$ is the relative momentum and $P=p_1+p_2$ is the total momentum of the mesons. 
The labels $n_{1,2}$ are assigned from the less energetic to the most energetic internal states at $P=0$.
Since we are treating the two mesons as non-interacting particles, we have some strong constraints on the allowed values of the relative momentum $p$. 
First, the finite-volume quantization of $p$ is affected by the reduced effective volume where the kinks can move freely due to the constraint that  
they cannot overlap.
Accordingly, the relative momentum is quantized as $p=m {\pi}/{L_{\rm eff}}$ with $m$ integer. 
The effective available volume is  $L_{\rm eff}=L-2$ for two interchain mesons (two kinks on each chain)
and $L_{\rm eff}=L-4$ for two intrachain mesons (the four kinks are all on the same chain).
Moreover, since they cannot overlap, the states with $p=0$ are forbidden. 
These two reasonable assumptions will be justified also a posteriori by the correct description of the relevant part of the energy spectrum.

The approximation of non-interacting mesons works for large enough $\Delta_{\perp}$ and in the limit $\Delta_{||}\gg 1$. 
Indeed, when $\Delta_{\perp}$ becomes too small, the internal oscillations of each meson become so wide that the 4-kink states cannot be interpreted as a composition of 
separate mesons.
Furthermore, this approximation is not expected to be effective for large $n_{1,2}$ because higher meson states have a more extended wave function 
(as it can be immediately deduced by looking at the spreading of the Bessel functions with respect to their index).

\begin{figure}[!t]
\centering
\includegraphics[width=0.8\textwidth]{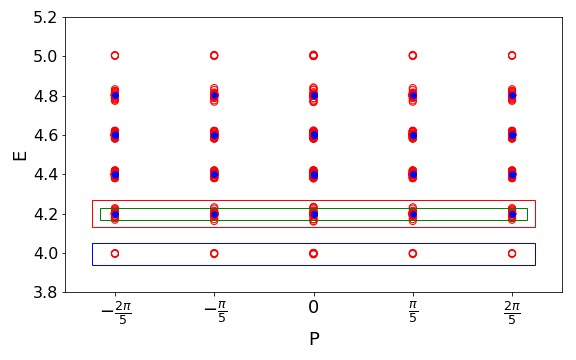}
\caption{
Spectrum of the ladder (red circles) and the staggered XXZ chain (blue dots) for energies in the interval $E\in[4,5]$ where four-kink states lie for large $\Delta_{||}$.   
We report data for the zero magnetization sector in both chains $M_1=M_2=0$. We work at $\Delta_\perp=5$, $\Delta_{||}=100$,  and $L=10$. 
The blue, red, and green rectangles are the regions that are magnified in Fig. \ref{fig:4kinksh5delta100-6}.
}
\label{fig:4kinksh5delta100}
\end{figure}

\begin{figure}[!t]
\centering
\includegraphics[width=1\textwidth]{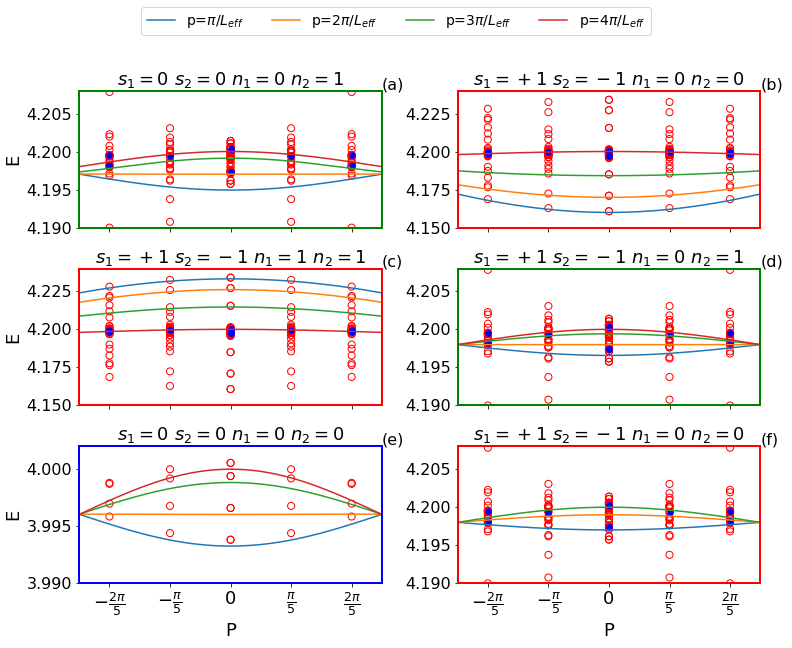}
\caption{
Zooms of the marked areas in Fig. \ref{fig:4kinksh5delta100}. The different symbols are the numerical data. 
The frame colors (red, green, and blue) correspond to the colors of the rectangles in Fig. \ref{fig:4kinksh5delta100}. 
The green area is a further zoom of the red one. 
The continuous lines correspond to the noninteracting two-meson approximation \eqref{eq:4kink} for interchain `Type 2' (a,b,c,d,e) and intrachain `Type1' (f) mesons. 
Each panel shows these approximate energy levels for fixed values of the spins and internal labels of the mesons corresponding to the quantum numbers reported above each panel. 
The different lines correspond to different values of the relative momentum $p$ reported in the legend on top of the plot. $L_{\rm eff}=L-2$ or $L_{\rm eff}=L-4$ for interchain 
and intrachain mesons, respectively.}
\label{fig:4kinksh5delta100-6}
\end{figure}

We compare the spectra of the ladder and of the staggered XXZ chain in Figs. \ref{fig:4kinksh5delta100} and \ref{fig:4kinksh5delta100-6}. 
In the former figure we report all states in the energy interval $[4,5]$ and identify some smaller windows (indicated by large rectangles) 
that are analyzed in detail in the latter figure. 
The first simple fact evident in both figures is that there are many more 4-kink states in the ladder than in the chain, reflecting the presence of interchain mesons
which do not exist on the chain. Hence, for the $4$-kink states, the mean-field treatment does not predict much.

We present a more quantitative analysis in Fig. \ref{fig:4kinksh5delta100-6}.
Here we zoom in the regions within the colored rectangles in Fig. \ref{fig:4kinksh5delta100}.
The spectrum in these windows is compared with the energy levels computed with Eq.~\erf{eq:4kink}. 
Notice that the red rectangle appears three times and the green one twice, where the green area is a further zoom of the red one. 
We make this choice because we plot the dispersion relation with different quantum numbers that cannot be put on the same graph 
in a clear manner.  
On top of each panel we report the quantum numbers corresponding to the spin and the internal energy levels of the mesons that are displayed in the plot.

Despite the roughness of the
approximation, it is remarkable that Eq.~\erf{eq:4kink}  captures some features of the spectrum. 
For example, the four dispersions plotted in the frames (b) and (c) very neatly describe families of states that are well separated from each other.
Also in panel (f), the intrachain states (that exist also for the staggered chain but with different momentum quantization, as it is 
clear from the fact that they alternate) are well captured by our approximation. 
Instead, resolving the states within the green frames in Figs. \ref{fig:4kinksh5delta100-6} is beyond the purpose of our approximation. 
It is a dense region where the separation of the states is comparable to higher perturbative orders in $\epsilon$ that are neglected in our description.
On the other hand, all the states that are sufficiently isolated in the spectrum (on an energy scale of order $\epsilon$), are well captured by this approximation. We remark that while, in virtue of confinement, the single-meson bands described in Sec. \ref{sec:exc} remain discrete and well separated in energy when $L$ goes to infinity, the two-meson bands become denser and fall in the continuum part of the spectrum.

\section{A transition for the first excited states}
\label{sec:nat}

\begin{figure}[t]
\centering
\includegraphics[scale=0.24]{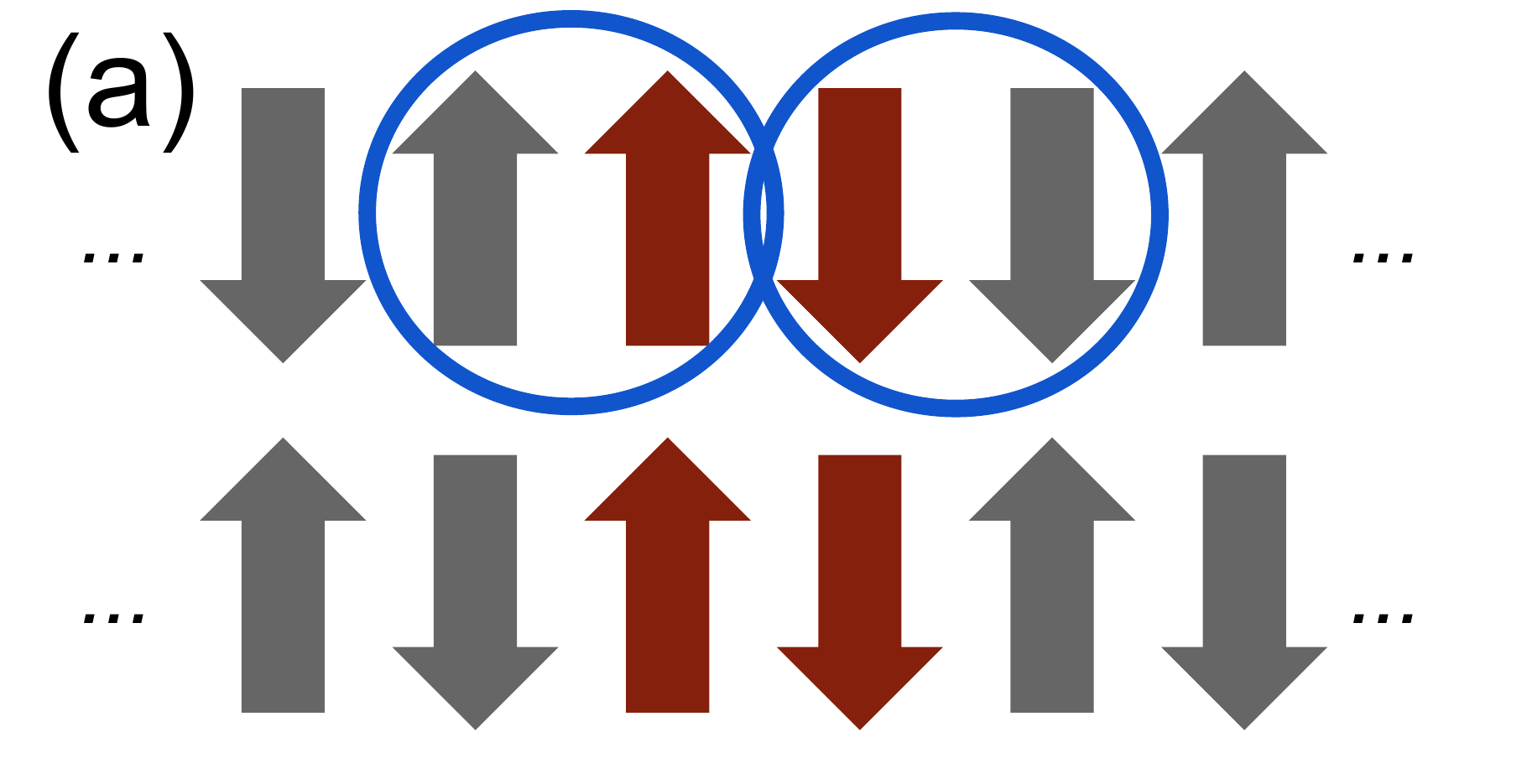}
\hfill
\includegraphics[scale=0.24]{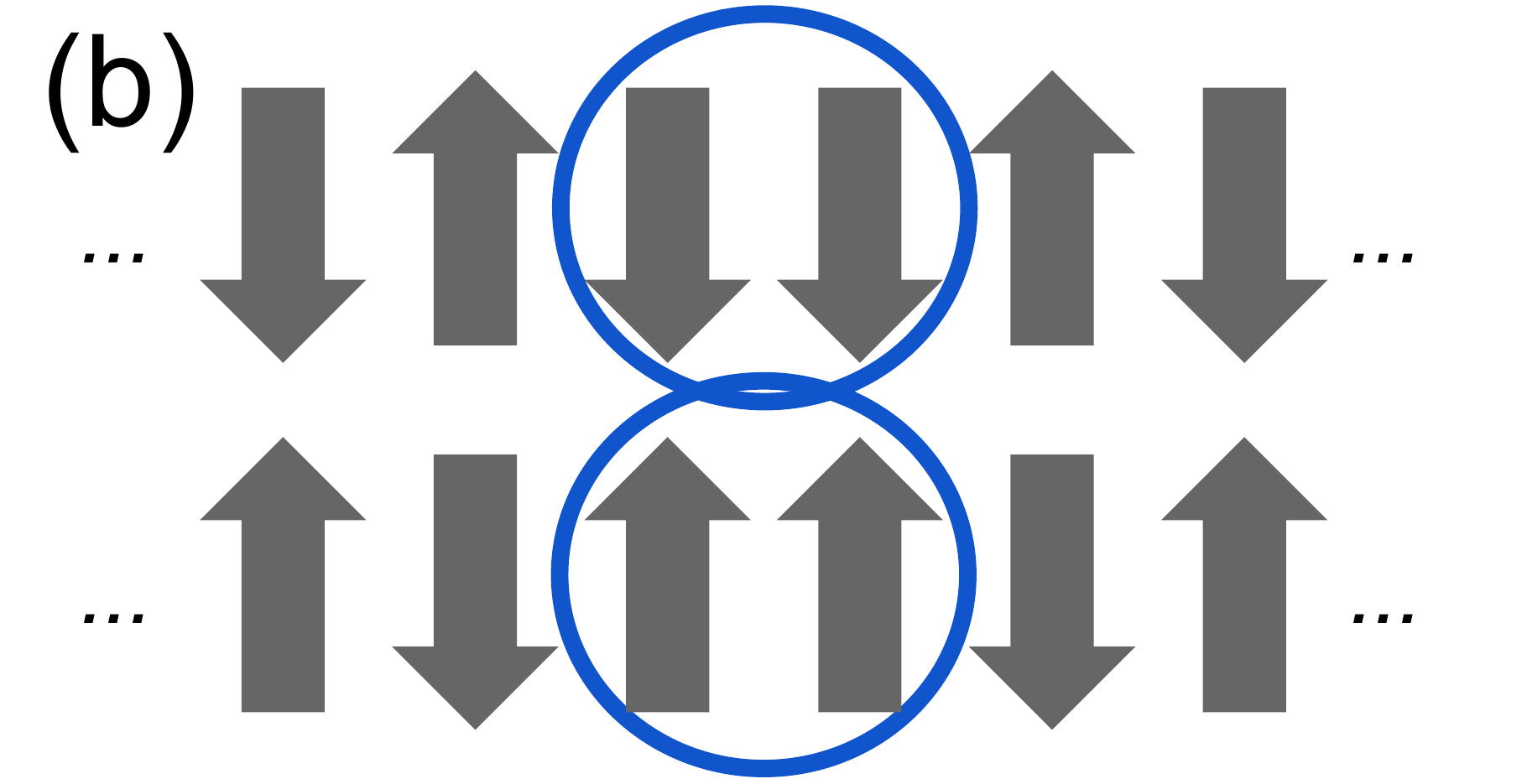}
\hfill
\includegraphics[scale=0.24]{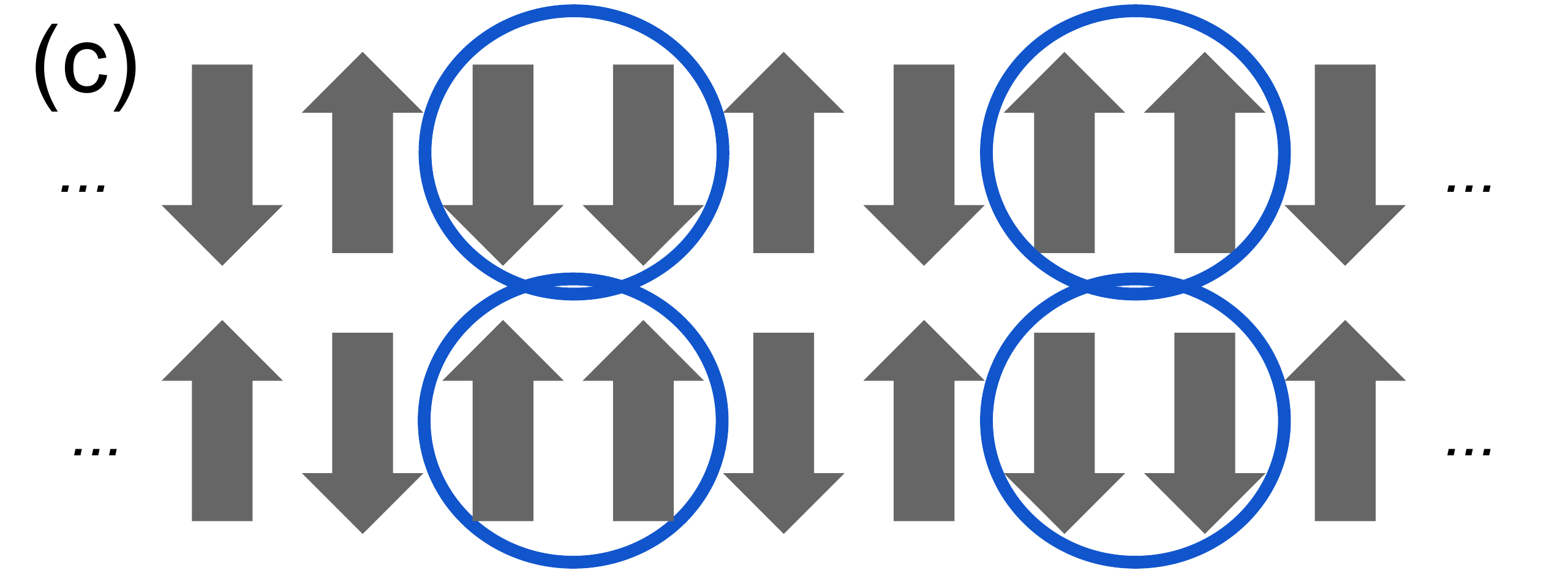}
\caption{Schematic pictures of the lowest-lying (a) intrachain and (b) interchain mesons. For $\Delta_{\perp}\lesssim 2\Delta_{||}$ the intrachain meson (a) represents the first excited state. In the regime $\Delta_{\perp}\gtrsim 2\Delta_{||}$ the low-energy sector is made of states with pairs of interchain mesons (c).
}
\label{fig:firstexc}
\end{figure}

We have already shown in Sec. \ref{sec:exc} that, in the limit of large $\Delta_{||}$ and moderate $\Delta_\perp$, the low-lying excitations of a ladder with even 
$L$ in the sector of zero magnetization  are well captured by an effective model of a spin chain in staggered field; 
the lowest excitations are intrachain (`Type 1') mesons, confined bound states of kinks.
Decreasing the value of $\Delta_{||}$, we observe that the nature of the first excited states changes qualitatively.
This can be understood from a simple classical argument.
The lowest lying intrachain meson (Fig.~\ref{fig:firstexc}-a) has energy $E_{\text{intra}} \simeq 2+4\Delta_{\perp}/\Delta_{||}$, while the least energetic interchain meson (Fig.~\ref{fig:firstexc}-b) has $E_{\text{inter}} \simeq 2$. 
Despite being less energetic, we cannot find this interchain excitation in the low-energy spectrum in the zero-magnetization sector, because, 
as discussed in Sec. \ref{sec:mesontype2}, a single meson of this type carries magnetizations $s_{1,2}=\pm 1/2$ on the two chains and it is only compatible with odd $L$. 
However, when $\Delta_{||}$ is sufficiently small, the energy of the first intrachain meson becomes so large that it is comparable with the energy $2E_{\text{inter}}\simeq 4$ 
of a two-meson state of `Type 2' (Fig.~\ref{fig:firstexc}-c). This happens when $2E_{\text{inter}}=E_{\rm intra}$, i.e.  for $\Delta_{||}\sim 2 \Delta_{\perp}$.

\begin{figure}[t]
\centering
\includegraphics[width=1\textwidth]{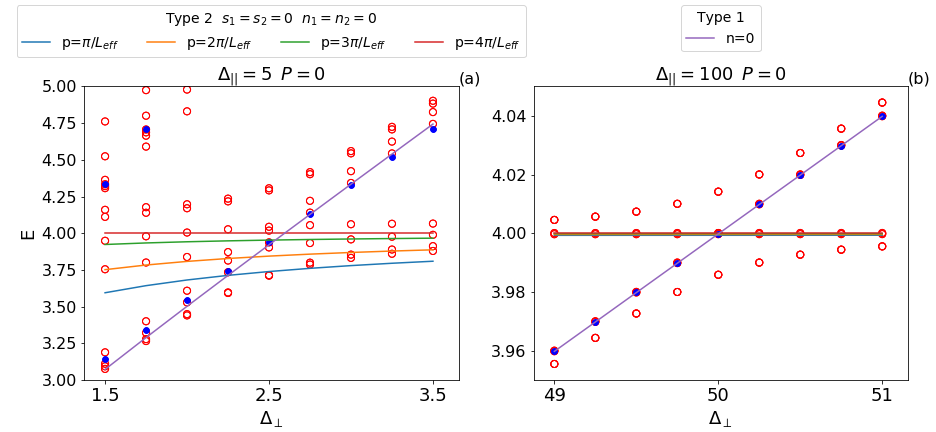}
\caption{Spectrum on the ladder (red circles) and on the staggered chain (blue dots) around the level crossing near $\Delta_{||}\approx 2 \Delta_{\perp}.$ The violet solid line shows the numerical prediction for the first level of intrachain mesons computed using Eq.~\eqref{eq:type1s0}. The other lines are extracted from Eq.~\eqref{eq:4kink}  for two interchain mesons with $s_1=s_2=0$ and $n_1=n_2=0$ like in Fig.~\ref{fig:4kinksh5delta100-6}-e. The corresponding quantum numbers are shown in the legend.}
\label{fig:crossing}
\end{figure}

We illustrate this transition in Fig. \ref{fig:crossing} for $\Delta_{||}=100$ and $\Delta_{||}=5$ by showing all the numerically calculated zero-momentum eigenstates in 
the relevant energy range as functions of $\Delta_\perp$, for $\Delta_\perp$ around $\Delta_{||}/2$. 
The blue dots represent the `Type 1' intrachain single meson states of the staggered XXZ chain.
As expected, when $\Delta_{\perp} \approx \Delta_{||}/2,$ the two kinds of states become nearly degenerate. At this point, the single meson and the 2-meson states hybridize.
 As the interchain two-meson states are invariant under the chain swap transformation, they only hybridize with intrachain meson states that are also invariant under chain swapping.

The observed phenomenon is not a quantum phase transition as it does not concern the ground state but the excited states. As a matter of fact, similar ``transitions'' take place already for smaller $\Delta_\perp$ at higher energy levels involving states with more kinks and mesons.
Nonetheless, the change in the nature of the low-energy sector has important physical consequences. It can be observed, for example, in the non-equilibrium dynamics after a quantum quench, where the spreading of correlations is determined by the quasiparticle excitations. 
While in the absence of confinement excitations can propagate freely, in the presence of an attractive potential quasiparticles get confined into mesons and hence the spreading of entanglement and correlations is suppressed. This is what we expect when $\Delta_{||}\gtrsim 2\Delta_{||}$. Conversely, for $\Delta_{||}\lesssim 2\Delta_{\perp}$, the low-energy sector is a continuum of two-meson states: while kinks are still confined in mesons, the spreading of correlation is not suppressed because the pairs of mesons are free to move with opposite momenta. Therefore, in a quench a dramatic difference between the two regimes is likely to emerge.

\section{Conclusions}
\label{concl}

In this manuscript we systematically characterized the spectrum of the Heisenberg-Ising ladder with Hamiltonian \eqref{eq:ladder} in the region
of parameters presenting confinement, i.e. in the ordered antiferromagnetic phase of the two chains for $\Delta_{||}>1$. 
Our main result is that we find two kinds of quasiparticle excitations, which we dub intrachain and interchain mesons, 
that correspond to bound states of kinks within the same chain or between different ones, respectively. 
Very importantly, intrachain mesons can be also obtained by means of a mean field treatment mapping the Hamiltonian to a staggered chain.  
Interchain mesons are genuine features of the ladder and they were not known by other means. 
They are expected to be a common characteristic of ladders with Ising-like rung interactions that lead to confinement. 
In fact, their existence is a consequence of the spontaneous braking of one symmetry. Hence, there are two 
equivalent true vacua and neutral mesons can interpolate between the same or different ones.  
One-particle intrachain (interchain) mesons are present only when the total ladder has even (odd) length. 
Conversely, two-particle states of interchain mesons are present also for even $L$. 

We quantitatively characterize the meson states. First, in the limit of large $\Delta_{||}$ we find the one-particle meson dispersion 
by projecting on the two-kink subspace. We release the condition of very large anisotropy exploiting semiclassical quantization. 
We also describe the four-kink (two mesons) states in the dilute approximation, i.e. treating the two mesons as non-interacting particles. 
Finally, we point out an interesting transition for the first excited state in even length ladders.  
At fixed $\Delta_{||}$, the first excited state is a one-particle intrachain meson state for small $\Delta_\perp$, but as the latter is increased 
it crosses over to a two interchain meson state. 

We finally discuss some lines of future research. 
A first question concerns the physics of more than two coupled chains (e.g. three for a start). 
Are there new kinds of bound states that can emerge from the enlarged local Hilbert space? 
It would be interesting to investigate also the case of
anisotropic Heisenberg-like (XXZ) interchain coupling which is more relevant to experiments on spin-chain compounds. Another interesting question concerns the influence of these bound states on the non-equilibrium quench dynamics, a subject that, as mentioned in the introduction, is nowadays under intense scrutiny. We plan to address this interesting topic in a future work.

\section*{Acknowledgments} 
We are extremely grateful to Sergei Rutkevich for discussions and valuables comments on a first version of this manuscript.
We thank G\'abor Tak\'acs for useful discussions and for granting access to the cluster at BME where the numerical calculations have been carried out. G.L. thanks BME for hospitality.
P.C. and G.L. acknowledge support from ERC under Consolidator grant  number 771536 (NEMO). F.M.S. acknowledges partial support from ERC under grant number 758329 (AGEnTh).
M. K. thanks SISSA for hospitality. M. K. acknowledges partial support from the National Research Development and Innovation Office (NKFIH) through Hungarian Quantum Technology National Excellence Program, project no. 2017-1.2.1-NKP-2017- 00001, through a K-2016 grant no. 119204 as well as grant no. SNN 118028, and from the BME-Nanotechnology FIKP grant of EMMI (BME FIKP-NAT).  M.K. was also supported by a “Bolyai J\'anos” grant of the HAS, and by the \'UNKP-19-4 New National Excellence Program of the Ministry for Innovation and Technology.

\section*{Note added} 
After completion of this work, we became aware of a work by F. B. Ramos {\it  et al.} \cite{new1} who independently developed the idea of intrachain and interchain mesons for a different ladder system.

\appendix

\section{Kink scattering phases}
\label{app:BA}

In this Appendix, we collect the exact expressions for the scattering phases which are used in the semiclassical quantization equations in Sec. \ref{sec:BS}.

In the gapped antiferromagnetic phase of the XXZ chain, in the absence of external magnetic fields, the elementary excitations are spin-$1/2$ topological excitations, $|K_{\alpha\beta}(\th)\rangle_s$ interpolating between the two degenerate vacua $\alpha,\beta.$ Their momenta and $z$ spin component are labeled by $\th$ and $s,$ respectively. Their dispersion relation can be parameterized by the so-called rapidity variable $\lambda\in[\pi/2,\pi/2]$ as \cite{Johnson1973}
\begin{subequations}
\label{eq:disprel}
\begin{align}
\th(\lambda)&=\frac\pi2-\mathrm{am}(2K\lambda/\pi,k)\,,\label{eq:thl}\\
\omega(\lambda)&=\frac{2K}\pi\sinh(\gamma)\mathrm{dn}(2K\lambda/\pi,k)\,,
\end{align}
\end{subequations}
where $K=K(k)$ is the complete elliptic integral of modulus $k$ with
\begin{equation}
\frac{K(\sqrt{1-k^2})}{K(k)}=\frac\gamma\pi\,,
\end{equation}
and $\mathrm{am}(x,k)$ and $\mathrm{dn}(x,k)$ are the Jacobi amplitude and delta amplitude. The parameterization \eqref{eq:disprel} is equivalent to the 
form \eqref{eq:disp} in the main text.

These particles are interacting which is manifested in their nontrivial scattering properties. In the total spin zero channel, corresponding to the scattering of a $s=1/2$ and a $s=-1/2$ particle, the scattering matrix is diagonalized by the combinations
\begin{equation}
|K_{\alpha\beta}(\th_1)K_{\beta\alpha}(\th_2)\rangle_{\pm} = \frac1{\sqrt{2}}\left(|K_{\alpha\beta}(\th_1)K_{\beta\alpha}(\th_2)\rangle_{\frac12,-\frac12}\pm|K_{\alpha\beta}(\th_1)K_{\beta\alpha}(\th_2)\rangle_{-\frac12,\frac12}\right)\,.
\end{equation}
The scattering phases are defined as
\begin{subequations}
\begin{align}
|K_{\alpha\beta}(\th_1)K_{\beta\alpha}(\th_2)\rangle_{ss} &= w_0(\th_1,\th_2)|K_{\alpha\beta}(\th_2)K_{\beta\alpha}(\th_1)\rangle_{ss}\,,\\
|K_{\alpha\beta}(\th_1)K_{\beta\alpha}(\th_2)\rangle_\pm &= w_\pm(\th_1,\th_2)|K_{\alpha\beta}(\th_2)K_{\beta\alpha}(\th_1)\rangle_\pm\,.
\end{align}
\end{subequations}
They were obtained in Ref. \cite{zabrodin1992} using Bethe ansatz with the result
\begin{align}
w(\th_1,\th_2) &= - e^{i\phi_\eta(\th_1,\th_2)}\,,\\
\phi_\eta(\th_1,\th_2) &= \Phi_\eta(\lambda_1-\lambda_2)\,,\\
\Phi_0(\lambda)&=-\lambda-\sum_{n=1}^\infty\frac{e^{-n\gamma}\sin(2\lambda n)}{n\cosh(n\gamma)}\,,\\
\Phi_\pm(\lambda)&=\Phi_0(\lambda) + \chi_\pm(\lambda)\,,\\
\chi_+(\lambda) &= - i\ln\left(-\frac{\sin[(\lambda-i\gamma)/2]}{\sin[(\lambda-i\gamma)/2]}\right)\,,\\
\chi_-(\lambda) &= - i\ln\left(\frac{\cos[(\lambda-i\gamma)/2]}{\cos[(\lambda-i\gamma)/2]}\right)
\end{align}
with $\th_k=\th(\lambda_k)$ as in Eq. \eqref{eq:thl}. The scattering phases $\phi_\eta(\th_1,\th_2)$ are the ones that appear in Eqs. \eqref{eq:BS} and \eqref{eq:BSL} with $\eta$ chosen according to the the total spin of the particles.

\end{document}